\documentclass[lettersize,journal]{IEEEtran}
\usepackage{amsmath,amsfonts}
\usepackage{algorithmic}
\usepackage{algorithm}
\usepackage{array}
\usepackage[caption=false,font=normalsize,labelfont=sf,textfont=sf]{subfig}
\usepackage{textcomp}
\usepackage{stfloats}
\usepackage{url}
\usepackage{verbatim}
\usepackage{graphicx}
\usepackage{cite}
\usepackage{array}
\usepackage{hyperref}
\usepackage{cleveref}
\usepackage{booktabs}
\usepackage{multirow}
\usepackage[T1]{fontenc}

\newcolumntype{C}[1]{>{\centering\arraybackslash}p{#1}}

\begin{document}

\title{EUR/USD Exchange Rate Forecasting incorporating Text Mining Based on Pre-trained Language Models and Deep Learning Methods}

\author{Hongcheng Ding, Xiangyu Shi, Ruiting Deng, Salaar Faroog, Deshinta Arrova Dewi\\ Bahiah A Malek, Shamsul Nahar Abdullah* 

~\IEEEmembership{INTI International University, Malaysia}

~\IEEEmembership{E-mail: shamsuln.abdullah@newinti.edu.my
\\ Orcid: 0000-0003-3242-0686}

\thanks{This paper was produced by the INTI International University}
\thanks{Manuscript finished October 19, 2024.}
\thanks{Email: i24025877@student.newinti.edu.my (H. Ding); i24026165@student .newinti.edu.my (F. Hu)}}



\maketitle

\begin{abstract}
This study introduces a novel approach for EUR/USD exchange rate forecasting that integrates deep learning, textual analysis, and particle swarm optimization (PSO). By incorporating online news and analysis texts as qualitative data, the proposed PSO-LSTM model demonstrates superior performance compared to traditional econometric and machine learning models. The research employs advanced text mining techniques, including sentiment analysis using the RoBERTa-Large model and topic modeling with LDA. Empirical findings underscore the significant advantage of incorporating textual data, with the PSO-LSTM model outperforming benchmark models such as SVM, SVR, ARIMA, and GARCH. Ablation experiments reveal the contribution of each textual data category to the overall forecasting performance. The study highlights the transformative potential of artificial intelligence in finance and paves the way for future research in real-time forecasting and the integration of alternative data sources.
\end{abstract}

\begin{IEEEkeywords}
EUR/USD exchange rate forecasting, deep learning, textual analysis, particle swarm optimization, LSTM, RoBERTa-Large, topic modeling, sentiment analysis
\end{IEEEkeywords}

\section{Introduction}
In the intricate tapestry of the global financial market, the exchange rate between the Euro and the US Dollar (EUR/USD) stands as a pivotal thread, weaving together the economic narratives of two powerhouse currencies. As a barometer of the economic relations between the Eurozone and the United States, the EUR/USD exchange rate holds profound implications for international trade, cross-border investments, and the overall health of the global economy. Accurately forecasting this exchange rate has become a holy grail for financial market participants, as it can provide a strategic edge in navigating the complex and ever-shifting currents of the foreign exchange market \cite{boyoukliev2022modelling, tsuji2022exchange}.

Recent advancements in the field of EUR/USD exchange rate forecasting have been propelled by the advent of machine learning and the proliferation of big data. State-of-the-art methodologies have harnessed the power of deep learning architectures, such as Long Short-Term Memory (LSTM) networks, to capture the non-linear and temporal dependencies inherent in financial time series data \cite{sarmas2022comparison, bejger2021forecasting}. These models have shown promising results in predicting exchange rate movements by learning from vast amounts of historical data and uncovering hidden patterns that traditional econometric models might overlook \cite{nemavhola2021application}.

However, despite the significant strides made by these cutting-edge approaches, there remain notable limitations. One critical drawback is the reliance on structured, quantitative data, such as historical prices and economic indicators \cite{kaushik2020forecasting}. While these data points provide valuable insights, they fail to fully capture the rich tapestry of qualitative information that shapes market sentiment and drives currency fluctuations. News articles, financial reports, and social media discussions often contain nuanced and real-time insights that can significantly influence exchange rate dynamics. Neglecting these qualitative data sources can lead to an incomplete understanding of the market and suboptimal forecasting performance.

To address this challenge, we propose a novel approach that seamlessly integrates both quantitative and qualitative data to enhance the accuracy and robustness of EUR/USD exchange rate forecasting \cite{sun2020ensemble, escudero2021neural}. Our methodology leverages the power of LLMs, specifically GPT-4, to process and extract relevant information from vast amounts of textual data. By employing advanced techniques such as prompt engineering and sentiment analysis, we enrich the predictive power of LSTM networks, enabling them to consider both historical price patterns and real-time market sentiments \cite{shen2021hybrid, nemavhola2021lstm}. Furthermore, we incorporate PPSO to fine-tune the hyperparameters of our model, ensuring optimal performance and adaptability to changing market conditions \cite{cao2020deep}.

This research introduces several innovative contributions to the field of financial forecasting, enhancing existing methodologies and introducing novel approaches:
\begin{itemize}
    \item We demonstrate how to effectively integrate state-of-the-art LLMs for preprocessing and annotating financial data, significantly improving the quality and relevance of the dataset for forecasting purposes. This involves utilizing LLMs and prompt engineering techniques to filter noise from high-noise news sources, ensuring cleaner and more accurate data for analysis.
    \item Our study showcases the combined use of twitter-RoBERTa-Large-topic-sentiment-latest and RoBERTa-Large models for extracting sentiment and identifying hidden relationships within textual data, thereby deepening and expanding the analytical scope.
    \item We introduce an innovative approach to EUR/USD exchange rate forecasting by employing a PSO-LSTM model, a concept that is relatively novel in this domain. This method signifies a shift from traditional econometric models to a comprehensive framework that integrates qualitative and quantitative data for improved forecasting accuracy.
\end{itemize}

The remainder of this paper is structured as follows: Section 2 provides a comprehensive literature review, positioning our research within the existing body of knowledge and highlighting the gaps we aim to address. Section 3 delves into the methodology and system design of our approach, including a detailed explanation of prompt engineering, LLMs, transformer architecture, and the PSO-LSTM model. Section 4 presents our empirical results, showcasing the performance of our model in comparison to baseline methods. Section 5 discusses the implications of our findings and conducts ablation studies to validate the contribution of each component of our approach. Finally, Section 6 concludes the paper, summarizing our key contributions, limitations, and outlining future research directions.

\section{Literature Review}
The landscape of exchange rate forecasting undergoes a significant transformation, transitioning from a reliance on traditional econometric models to the adoption of advanced machine learning and deep learning techniques. In the early stages, this field relies heavily on conventional statistical models such as Autoregressive Moving Average (ARMA), Autoregressive Integrated Moving Average (ARIMA), and Generalized Autoregressive Conditional Heteroskedasticity (GARCH) \cite{meese1983empirical, bollerslev1986generalized}. These models aim to predict exchange rate movements by analyzing the inherent features of historical time series data. However, they often struggle to capture the complex, non-linear dynamics and non-stationary characteristics of exchange rate time series \cite{engel2005exchange, rossi2013exchange}. This limitation of traditional econometric models highlights the need for more sophisticated and adaptive methodologies capable of handling the intricate nature of financial data.

The advent of machine learning techniques marks a significant milestone in the evolution of exchange rate forecasting. Algorithms such as Support Vector Machines (SVM) and Random Forests gain prominence due to their ability to uncover hidden patterns and trends in complex datasets \cite{huang2005forecasting, krauss2017statistical}. These methods offer a level of flexibility and accuracy that often surpasses that of traditional statistical models. For instance, Yu et al. \cite{yu2007forecasting} demonstrate the superior performance of SVM in forecasting exchange rates compared to conventional ARIMA models. Similarly, other researchers \cite{ozturk2016modeling} showcase the effectiveness of Random Forests in capturing the non-linear relationships in exchange rate dynamics. The success of these early machine learning applications paves the way for more advanced techniques and sets the stage for the integration of deep learning in exchange rate forecasting.

Despite the promising results of machine learning methods, the prediction of EUR/USD exchange rates continues to face challenges due to the limitations of relying solely on historical data. Traditional econometric models and early machine learning approaches primarily focus on analyzing past exchange rate trends, which often fail to capture the real-time dynamics and sentiment-driven fluctuations of the market. However, the emergence of big data, especially in the form of unstructured textual data, provides a new source of information that could potentially revolutionize exchange rate forecasting \cite{nassirtoussi2014text}. Online text mining and sentiment analysis become prominent areas of research, with studies demonstrating the value of incorporating market sentiment derived from various sources such as news articles, social media, and financial reports.

One notable example is the work of Das and Chen \cite{das2007yahoo}, who develop a method to extract investor sentiments from online forums and social media to predict stock market trends. Their approach showcases the potential of leveraging user-generated content to gauge market sentiment and improve prediction accuracy. Similarly, Tetlock \cite{tetlock2007giving} quantifies media sentiment by analyzing content from The Wall Street Journal and explores its impact on the stock market and EUR/USD exchange rate fluctuations. These studies highlight the importance of considering sentiment trends and topic-specific discussions in understanding and predicting exchange rate movements.

The introduction of deep learning technologies marks a significant leap forward in the field of exchange rate forecasting. Models such as LSTMnetworks and Gated Recurrent Units (GRU) demonstrate remarkable abilities in capturing intricate patterns and long-term dependencies in time series data \cite{hochreiter1997long, cho2014learning}. The multi-layered, non-linear architecture of these models allows them to learn complex representations from vast amounts of financial data, making them particularly well-suited for analyzing high-dimensional datasets \cite{lecun2015deep}. The application of deep learning in exchange rate forecasting opens up new avenues for research and showcases the potential for more accurate and robust predictions.

However, the incorporation of textual information into deep learning models for exchange rate prediction remains relatively unexplored. While studies like Smales \cite{smales2014news} and Beckmann et al. \cite{beckmann2015exchange} make early attempts to integrate news sentiment into financial market forecasting, the full potential of leveraging large-scale textual data through advanced language models has yet to be realized. The development of transformer-based models, such as BERT \cite{devlin2018bert}, opens up new possibilities for mining textual information and extracting valuable insights for exchange rate forecasting. These models, pre-trained on vast amounts of unlabeled text data, capture intricate semantic relationships and contextual information, making them ideally suited for processing financial news and sentiment analysis.

Recent advancements in natural language processing (NLP) further expand the horizons of exchange rate forecasting by enabling the fusion of heterogeneous data sources. Researchers begin to explore the integration of structured data, such as macroeconomic indicators, with unstructured data, including news articles and social media posts. For instance, Serrano-Cinca et al. \cite{serrano2019using} propose a framework that combines traditional financial data with sentiment extracted from Twitter to predict stock market movements. Similarly, Pascal et al. \cite{spalti2022exploiting} investigate the use of satellite imagery data as an auxiliary information source to enhance exchange rate forecasting. These studies underscore the growing trend of multi-source data fusion and highlight the potential benefits of incorporating diverse data types to capture a more comprehensive view of the factors influencing exchange rates.

Furthermore, the rise of large language models, such as GPT-3 \cite{brown2020language}, unlocks new possibilities for processing and understanding vast amounts of unstructured text data. These models, with their unprecedented capacity for language understanding and generation, have the potential to revolutionize sentiment analysis and information extraction from financial texts \cite{pornwattanavichai2022bertforex, pfahler2021advanced}. By leveraging the power of large language models, researchers develop more sophisticated techniques for mining textual data and uncovering hidden patterns and relationships that can inform exchange rate predictions.

The field of exchange rate forecasting witnesses a significant evolution, transitioning from traditional econometric models to advanced machine learning and deep learning techniques. The limitations of conventional methods in capturing the complex dynamics of exchange rates drive the adoption of more sophisticated approaches \cite{tak2022foreign, shang2023datasets}. The incorporation of unstructured textual data and sentiment analysis emerges as a promising direction, with studies demonstrating the value of considering market sentiment and topic-specific discussions in predicting exchange rate movements \cite{lin2022deep, claveria2022deep}. The advent of deep learning, particularly with the integration of transformer-based models and large language models, opens up new avenues for leveraging large-scale textual data and extracting valuable insights. Moreover, the fusion of heterogeneous data sources, including structured and unstructured data, becomes an emerging trend in exchange rate forecasting research \cite{argotty2023novel}.

\section{Methodology}

\subsection{System Design}

\begin{figure*}[!ht]
    \centering
    \includegraphics[width=1\linewidth]{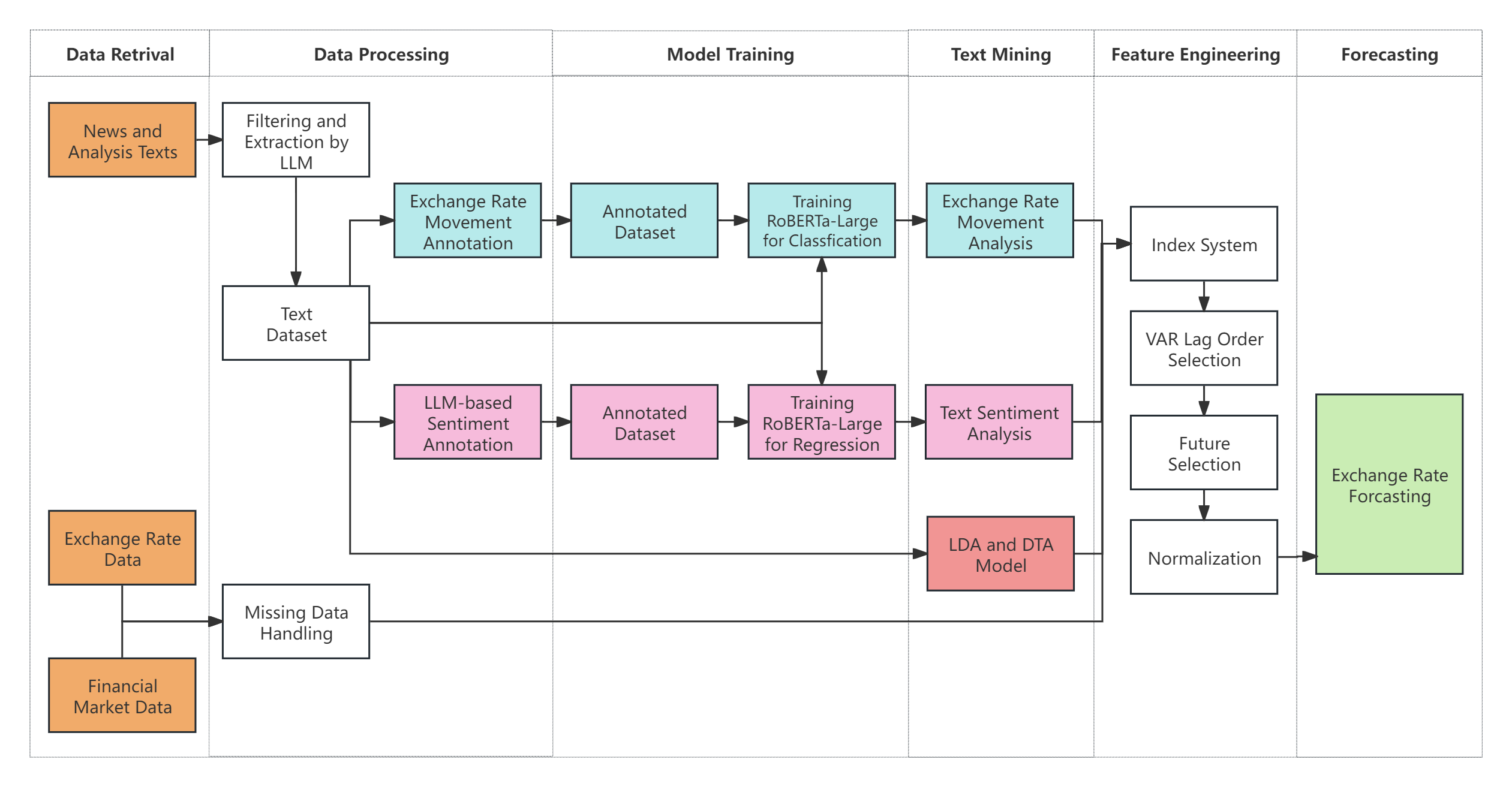}
    \caption{The framework of this work.}
    \label{fig:framework}
\end{figure*}

Our research design seamlessly integrates advanced text analysis techniques with robust financial modeling to forecast exchange rate movements accurately, shown in Figure \ref{fig:framework}. The process begins with the collection and preprocessing of relevant data, including exchange rates, financial news, and market data. The preprocessing stage involves normalization and sentiment annotation using a fine-tuned BERT model, ensuring data quality and consistency.

Following the preprocessing, we conduct sentiment analysis to extract market sentiments from news texts. These sentiment features are then combined with key financial indicators to create a comprehensive feature set that will inform our forecasting models. The careful selection and integration of these features are critical to capturing the complex dynamics of the foreign exchange market.

In the final phase, the curated features are fed into a Vector Autoregression (VAR) model to predict EUR/USD exchange rate movements. The VAR model, a mainstay in econometric modeling, is well-suited to capture the interdependencies among multiple time series variables. By incorporating both quantitative financial data and qualitative sentiment features, our VAR model aims to provide a holistic view of the factors driving exchange rate fluctuations.

The effectiveness of our forecasting model is evaluated by comparing its predictions against actual market behavior. Particular emphasis is placed on assessing the added value of textual sentiment analysis in capturing market dynamics. This evaluation process provides insights into the model's performance and guides further refinements to enhance its predictive accuracy.

\subsection{Data Processing for Textual Data}
\begin{figure}[!ht]
\centering
\includegraphics[width=1\linewidth]{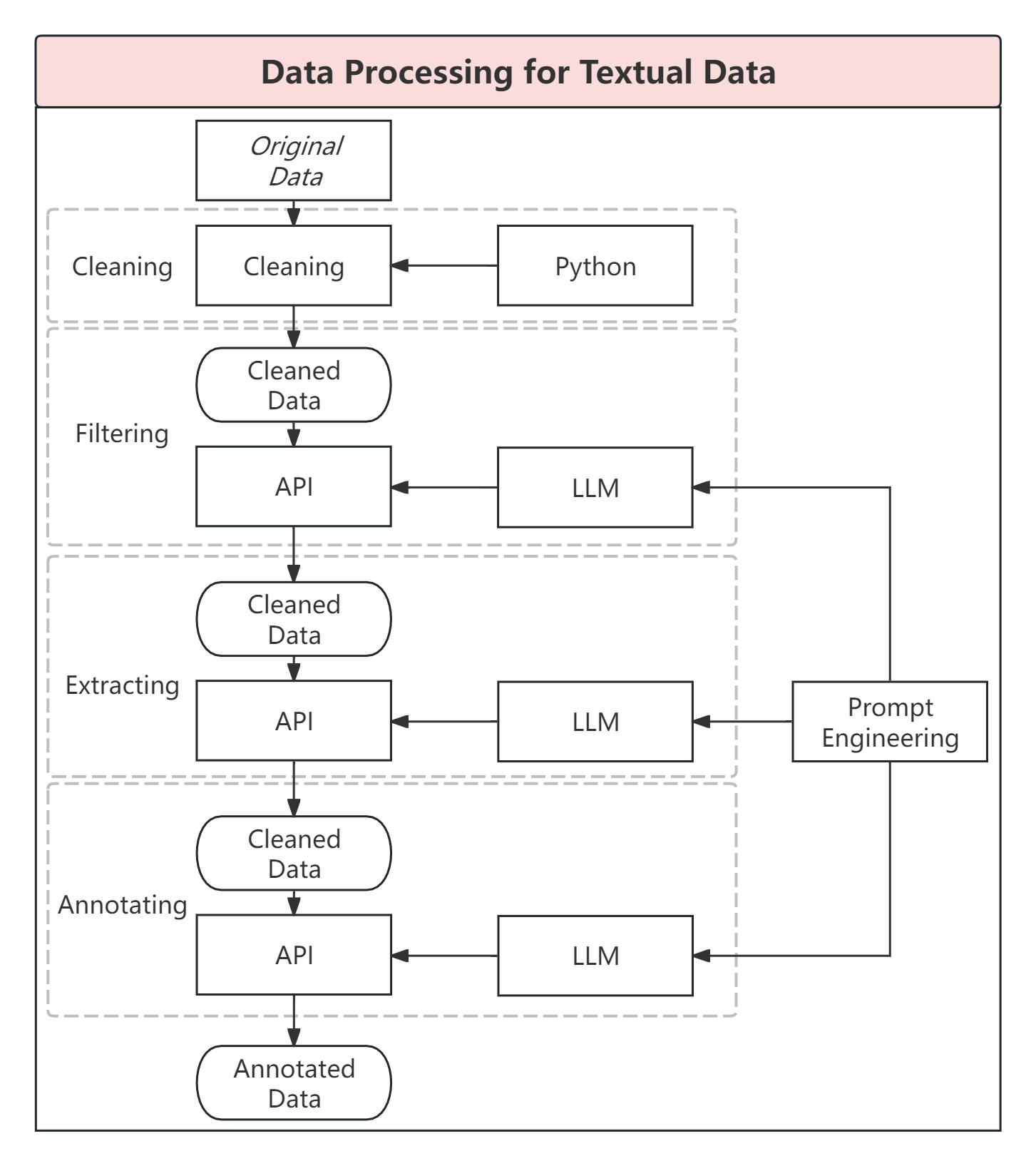}
\caption{Data processing framework for textual data using LLMs and prompt engineering.}
\label{fig:data processing}
\end{figure}

Our methodology for processing textual data involves a multi-stage approach, as illustrated in Figure \ref{fig:data processing}. The process begins with the original data, which undergoes a series of transformations to extract valuable insights.

The first stage is data cleaning, where the raw data is processed using Python scripts to remove noise, inconsistencies, and irrelevant information. The cleaned data is then passed through an API, which serves as an interface between the cleaning stage and the subsequent filtering stage.

In the filtering stage, the cleaned data is further refined using an LLM. The LLM applies advanced natural language processing techniques to filter out unwanted elements and retain only the most pertinent information. This filtered data is then passed to the extracting stage via another API.

The extracting stage employs the same LLM to extract key features, patterns, and insights from the filtered data. The LLM's ability to understand context and semantics enables it to identify the most salient aspects of the text. The extracted information is then channeled through an API to the final annotating stage.

In the annotating stage, the LLM is utilized once more to annotate the extracted data with meaningful labels, tags, or categories. This annotation process enriches the data by adding structured metadata, making it more accessible and interpretable for further analysis.

Throughout the data processing pipeline, prompt engineering plays a crucial role in guiding the LLM's behavior and outputs. Carefully designed prompts are used to elicit precise, contextually relevant, and coherent responses from the model at each stage. This targeted approach ensures that the LLM focuses on the specific aspects of the task at hand, yielding results that align closely with the desired objectives.

By leveraging the power of LLMs and prompt engineering, our methodology enables the efficient and effective processing of textual data, transforming raw information into valuable insights that can inform decision-making and analysis in various domains.

\subsection{LLMs}

LLMs like CHATGPT-4.0 have revolutionized the field of Natural Language Processing (NLP). These models, built upon deep learning architectures and trained on vast amounts of data, possess remarkable capabilities in understanding and generating human language \cite{Bommasani2021}.

The core strength of LLMs lies in their ability to capture the intricate structures and semantic patterns of language. By learning from extensive text corpora, LLMs develop a deep understanding of vocabulary, grammar, context, and the nuances of language use \cite{Brown2020}. This enables them to generate fluent, coherent, and contextually relevant text, making them highly adaptable to various linguistic settings.

One of the key advantages of LLMs is their ability to understand context and infer implicit meanings. Rather than merely processing surface-level information, LLMs can grasp the underlying intent, emotional tone, and contextual connections within the text \cite{Devlin2019}. This depth of understanding stems from their vast parameter space and complex neural network architectures, enabling them to handle ambiguous or context-dependent language with greater accuracy.

Moreover, the continuous training process of LLMs on diverse datasets equips them with broad cross-cultural and cross-domain knowledge. This makes them particularly powerful in applications that span multiple languages, cultures, and subject areas. The versatility of LLMs has opened up new possibilities in NLP, pushing the boundaries of what artificial intelligence can achieve in language understanding and generation.

\begin{figure}[!ht]
    \centering
    \includegraphics[width=1\linewidth]{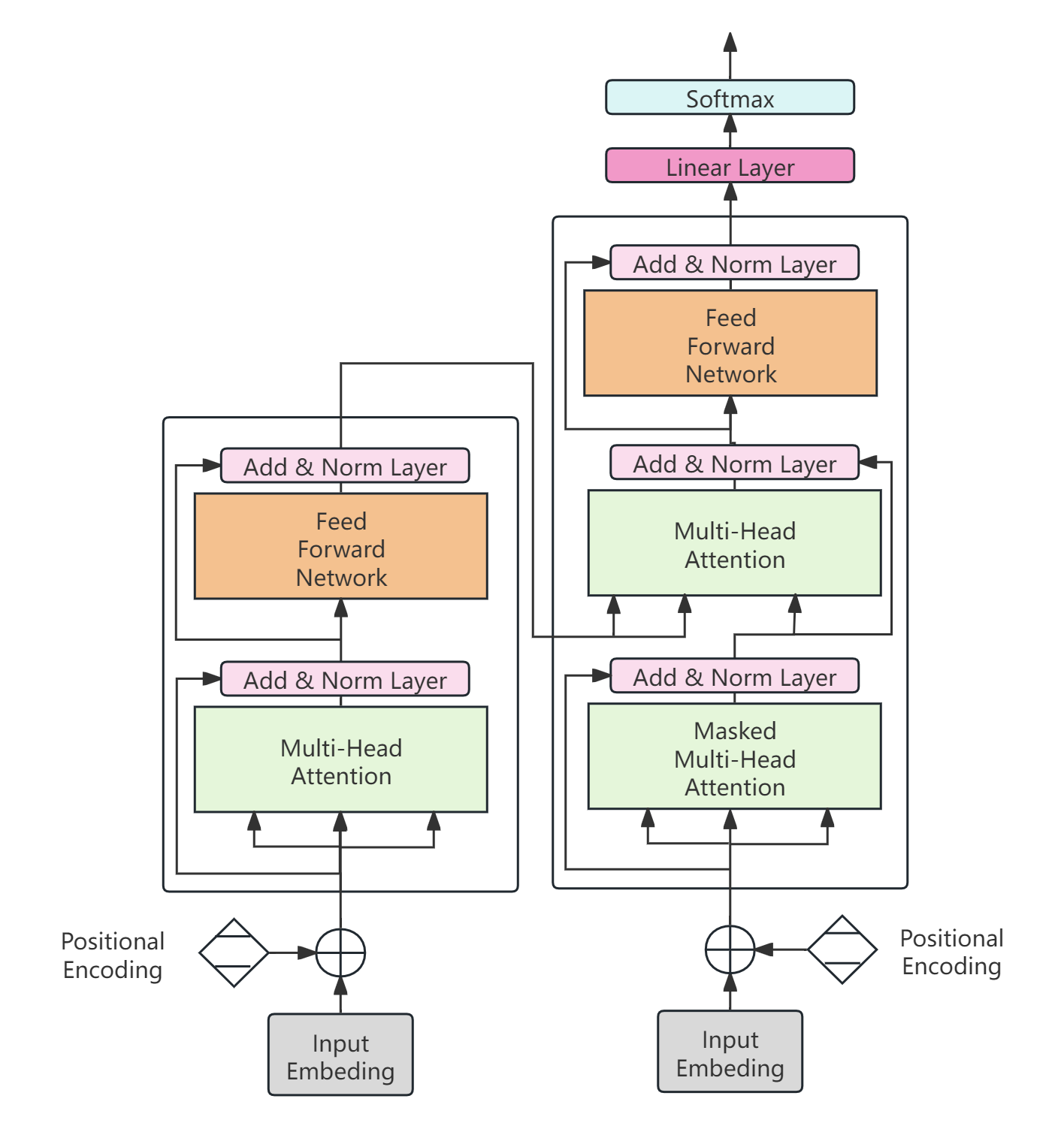}
    \caption{Transformer architrecture.}
    \label{fig:transformer}
\end{figure}

The CHATGPT-4.0 and BERT models, both based on the transformer architecture shown in Figure \ref{fig:transformer}, exemplify the state-of-the-art in LLMs. In their encoder components, they process input text through stacked multi-head self-attention mechanisms and fully connected networks \cite{Vaswani2017}. The input text is represented as a sum of token embeddings, segment embeddings, and positional embeddings, capturing both the semantic and positional information.

The encoder consists of multiple layers, each containing a multi-head self-attention sublayer and a feed-forward network (FFN) sublayer. Residual connections and layer normalization are applied between the sublayers to facilitate gradient flow and stabilize training.

The attention mechanism, a core component of transformers, enables the model to weigh the relevance of different input tokens dynamically. By computing attention scores between query, key, and value vectors, the model can focus on the most informative parts of the input sequence. The multi-head attention mechanism further enhances this by allowing the model to attend to information from different representation subspaces simultaneously.

\begin{equation}
\text{Attention}(Q, K, V) = \text{softmax}\left(\frac{QK^T}{\sqrt{d_k}}\right)V
\end{equation}
where \( Q \), \( K \), and \( V \) are the query, key, and value matrices, and \( d_k \) is the dimension of the key vectors.

The attention scores are computed by taking the dot product of the query and key matrices, scaled by the square root of the key dimension to prevent vanishing gradients. The scores are then normalized using the softmax function and used to weight the value matrix, yielding the final attention output.

Multi-head attention allows the model to attend to information from different representation subspaces in parallel. The outputs from multiple attention heads are concatenated and linearly transformed to produce the final output.

\begin{equation}
\text{MultiHead}(Q, K, V) = \text{Concat}(\text{head}_1, \ldots, \text{head}_h)W^O
\end{equation}
where \( \text{head}_i = \text{Attention}(QW_i^Q, KW_i^K, VW_i^V) \).

The multi-head attention output is then passed through a position-wise feed-forward network, which applies two linear transformations with a ReLU activation in between.

\begin{equation}
\text{FFN}(x) = \max(0, xW_1 + b_1)W_2 + b_2
\end{equation}

The decoder follows a similar structure to the encoder, with an additional multi-head attention sublayer that attends to the encoder's output. This allows the decoder to incorporate information from the input sequence when generating the output.

The combination of attention mechanisms, feed-forward networks, and residual connections enables transformers to capture long-range dependencies and learn complex language patterns effectively.

\subsection{Sentiment Analysis with RoBERTa-Large}

For sentiment analysis, we employ the RoBERTa-Large model, a variant of the RoBERTa (Robustly Optimized BERT Pretraining Approach) model \cite{Liu2019}. RoBERTa builds upon the BERT architecture, introducing several key modifications to improve performance and robustness.

The RoBERTa-Large model is pre-trained on a vast corpus of unlabeled text, allowing it to develop a deep understanding of language structure and semantics. We then fine-tune the model on our specific sentiment analysis task, leveraging its pre-trained knowledge to achieve high accuracy in understanding the nuances of financial language.

One of the strengths of the RoBERTa-Large model is its bidirectional architecture, which enables it to capture context from both past and future tokens in the input sequence. This bidirectional understanding is particularly crucial in sentiment analysis, where the sentiment of a word or phrase can be heavily influenced by its surrounding context.

In our study, we fine-tune the RoBERTa-Large model to classify the sentiment of financial news articles based on the movement of the EUR/USD exchange rate within a single trading day. The model is trained to predict the sentiment label \( M_t \) based on the day's closing exchange rate \( P_t \):

\begin{equation}
M_t =
\begin{cases}
0, & P_t < P_{t-1} \\
1, & P_t \geq P_{t-1}
\end{cases}
\end{equation}

By learning to associate patterns in the text with the corresponding exchange rate movements, the RoBERTa-Large model can uncover latent relationships between sentiment and market dynamics.

The fine-tuning process involves careful selection of hyperparameters such as learning rate, batch size, and the number of training epochs. We employ a systematic approach to hyperparameter optimization, evaluating different configurations to identify the most effective settings. The details of our hyperparameter search are provided in Appendix B.
\subsection{Time Series Forecasting with LSTM}

To capture the temporal dependencies in the EUR/USD exchange rate time series, we employ the LSTMmodel \cite{Hochreiter1997}. LSTM is a variant of Recurrent Neural Networks (RNNs) designed to overcome the limitations of traditional RNNs, such as the vanishing and exploding gradient problems.

\begin{figure}[!ht]
    \centering
    \includegraphics[width=0.6\linewidth]{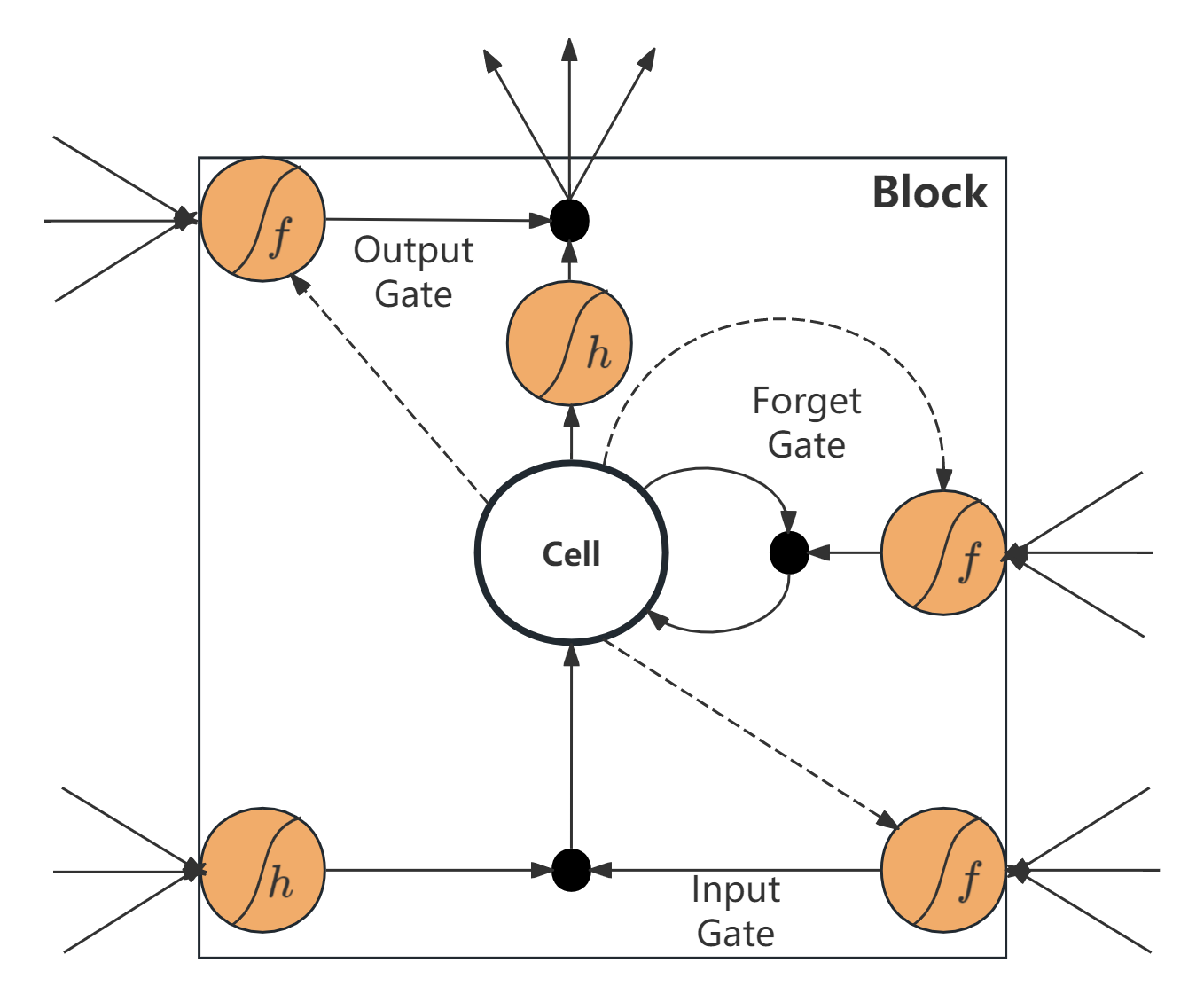}
    \caption{The mechanism of LSTM network.}
    \label{fig:LSTM}
\end{figure}

The key innovation of the LSTM architecture is the introduction of memory cells and gating mechanisms, shown in Figure \ref{fig:LSTM}. Each LSTM unit consists of an input gate, a forget gate, and an output gate, which regulate the flow of information into and out of the memory cell. This allows the model to selectively retain or forget information over long sequences, enabling it to capture both short-term and long-term dependencies.

The input gate \( i_t \) controls the amount of new information added to the memory cell, while the forget gate \( f_t \) determines what information should be discarded. The output gate \( o_t \) regulates the exposure of the memory cell to the next layer. The memory cell state \( c_t \) represents the internal state of the LSTM unit, which is updated based on the gating mechanisms.

The computations within an LSTM unit can be summarized as follows:

\begin{align}
f_t &= \sigma(W_f \cdot [h_{t-1}, x_t] + b_f) \\
i_t &= \sigma(W_i \cdot [h_{t-1}, x_t] + b_i) \\
C'_t &= \tanh(W_C \cdot [h_{t-1}, x_t] + b_C) \\
c_t &= f_t \cdot c_{t-1} + i_t \cdot C'_t \\
o_t &= \sigma(W_o \cdot [h_{t-1}, x_t] + b_o) \\
h_t &= o_t \cdot \tanh(c_t)
\end{align}

where \( W_f \), \( W_i \), \( W_o \), and \( W_C \) are weight matrices; \( b_f \), \( b_i \), \( b_o \), and \( b_C \) are bias terms; \( \sigma \) is the sigmoid activation function; and \( \tanh \) is the hyperbolic tangent activation function.

The LSTM model takes as input a sequence of historical exchange rates and outputs a prediction for the future exchange rate. The length of the input sequence, known as the time step length, determines the amount of historical data used to make each prediction. This allows the model to learn both short-term and long-term patterns in the exchange rate dynamics \cite{Graves2012}.

The number of LSTM units in the hidden layer determines the model's capacity to learn complex representations of the input data. These units are fully connected, processing the input through weighted summation and applying activation functions to produce the output.

By capturing the temporal dependencies and learning robust representations of the exchange rate time series, the LSTM model enables accurate forecasting of future EUR/USD exchange rate movements.

\subsection{Hyperparameter Optimization with PPSO}

To optimize the hyperparameters of our LSTM model, we employ the PPSO algorithm \cite{Kennedy1995}. PSO is a metaheuristic optimization technique inspired by the social behavior of bird flocks or fish schools.

In the context of hyperparameter optimization, each particle in the swarm represents a candidate solution, i.e., a set of hyperparameter values. The particles move through the search space, adjusting their positions based on their own best-known position (personal best) and the best position discovered by the entire swarm (global best).

The movement of each particle \( i \) at time step \( t \) is governed by the following equations:

\begin{equation}
V_{i}^{t+1} = w_i V_{i}^t + c_1 r_1 \left(P_{\text{best}}^i - X_{i}^t \right) + c_2 r_2 \left(P_{\text{gbest}} - X_{i}^t \right)
\end{equation}

\begin{equation}
X_{i}^{t+1} = X_{i}^t + V_{i}^{t+1}
\end{equation}

where \( V_{i}^t \) is the velocity of particle \( i \) at time \( t \), \( X_{i}^t \) is the position of particle \( i \) at time \( t \), \( w_i \) is the inertia weight, \( c_1 \) and \( c_2 \) are acceleration coefficients, \( r_1 \) and \( r_2 \) are random numbers between 0 and 1, \( P_{\text{best}}^i \) is the personal best position of particle \( i \), and \( P_{\text{gbest}} \) is the global best position of the swarm.

The objective of the optimization is to find the set of hyperparameters that minimizes the forecasting error of the LSTM model. We use the root mean square error (RMSE) as the fitness function to evaluate the quality of each candidate solution. The PSO algorithm iteratively updates the particles' positions until a predefined stopping criterion is met, such as a maximum number of iterations or a convergence threshold.

By leveraging the global search capabilities of PSO, we can efficiently explore the hyperparameter space and identify the optimal configuration for our LSTM model, leading to improved forecasting accuracy.

\subsection{Evaluation Metrics}

To assess the performance of our proposed PSO-LSTM model, we compare it against benchmark models such as GARCH and SVM. We employ two widely used evaluation metrics to quantify forecasting accuracy: MAE and RMSE \cite{Hyndman2006}.

The MAE measures the average absolute difference between the predicted and actual values:

\begin{equation}
\text{MAE} = \frac{1}{n} \sum_{i=1}^{n} |y_i - \hat{y}_i|
\end{equation}

The RMSE is the square root of the mean squared error and provides a measure of the average deviation of the predictions from the actual values:

\begin{equation}
\text{RMSE} = \sqrt{\frac{1}{n} \sum_{i=1}^{n} (y_i - \hat{y}_i)^2}
\end{equation}

where \( y_i \) is the actual value, \( \hat{y}_i \) is the predicted value, and \( n \) is the number of samples.

By evaluating our model using these metrics, we can comprehensively assess its forecasting performance and compare it against benchmark models. This evaluation process allows us to validate the effectiveness of our proposed approach and demonstrate its potential for accurate EUR/USD exchange rate prediction.

\section{Procedure of Experiment}

\subsection{Data Collection, Processing, and Cleaning}

We employ web scraping techniques to gather comprehensive datasets from \texttt{investing.com} and \texttt{forexempire.com}, covering the period from February 11, 2016, to January 19, 2024. The collected data undergoes preprocessing steps, including filtering, normalization, and summarization, to ensure its readiness for advanced text mining tasks. To handle the inherent noise in the original news data, the GPT-4.0 TURBO API \cite{radford2019language} is employed for data cleansing, leveraging its capabilities to retain essential information while removing irrelevant content. This process facilitates accurate subsequent analyses, such as Latent Dirichlet Allocation (LDA) \cite{blei2003latent}, Dynamic Topic Modeling (DTM) \cite{blei2006dynamic}, and sentiment analysis, by focusing the data on relevant content and enhancing the precision of thematic and sentiment extraction.

\begin{table*}[!ht]
\caption{Index system.}\label{tab:financial_indicators}
\renewcommand{\arraystretch}{0.8} 
\centering
\resizebox{\textwidth}{!}{ 
\begin{tabular}{p{4.5cm} p{1cm} p{4.2cm} p{2cm}} 
\toprule
\textbf{Classification} & \textbf{No.} & \textbf{Indicator Name} & \textbf{Reference} \\
\midrule
Target Series & 1 & EUR/USD Exchange Rate & \\
\midrule
US Exchange Rate & 2 & USD/CAD Exchange Rate & \cite{yang2024weighted,barunik2017asymmetric,greenwood2021measuring,kilic2017contagion} \\
(Top 5 Trading Partners) & 3 & USD/MXN Exchange Rate & \\
 & 4 & USD/CNY Exchange Rate & \\
 & 5 & USD/JPY Exchange Rate & \\
 & 6 & USD/KRW Exchange Rate & \\
\midrule
EU Exchange Rate & 7 & EUR/CNY Exchange Rate & \cite{yang2024weighted,barunik2017asymmetric,greenwood2021measuring,kilic2017contagion} \\
(Top 5 Trading Partners) & 8 & EUR/GBP Exchange Rate & \\
 & 9 & EUR/CHF Exchange Rate & \\
 & 10 & EUR/RUB Exchange Rate & \\
 & 11 & EUR/TRY Exchange Rate & \\
\midrule
Currency Index & 12 & US Dollar Index & \cite{alkan2022currency,ding2021conditional,chantarakasemchit2020forex,gurrib2016optimizing} \\
 & 13 & Euro Index & \cite{alkan2022currency,ding2021conditional,chantarakasemchit2020forex,gurrib2016optimizing} \\
\midrule
Currency Futures & 14 & US Dollar Futures-Jun & \cite{tornell2012speculation,ferraro2015can,chen2003commodity} \\
 & 15 & Euro Futures-Jun & \cite{lyons1997simultaneous} \\
\midrule
Commodities & 16 & Crude Oil WTI Futures & \cite{cashin2004commodity,jadidzadeh2017does} \\
 & 17 & Natural Gas Futures & \cite{lizardo2010oil,basher2016impact} \\
 & 18 & Gold Futures & \cite{pukthuanthong2011gold,sjaastad2008price} \\
 & 19 & Copper Futures & \cite{zhang2016exchange,ferraro2015can} \\
 & 20 & Corn Futures & \cite{nazlioglu2012oil,akanni2020returns,nazlioglu2013volatility,rezitis2015relationship} \\
 & 21 & Soybeans Futures & \cite{nazlioglu2012oil,akanni2020returns,nazlioglu2013volatility,rezitis2015relationship} \\
\midrule
Bond Yield & 22 & US 10-Year Bond Yield & \cite{engel2023forecasting,chinn2004monetary,lace2015determining} \\
 & 23 & Eurozone 10-Year Bond Yield & \cite{afonso2018euro,cecioni2018ecb} \\
\midrule
Interbank Offered Rate & 24 & SOFR - 1 month & \cite{tonzer2015cross,ivashina2015dollar,dal2012short,duffie2015reforming,du2018deviations} \\
 & 25 & EURIBOR - 1 month & \cite{tonzer2015cross,ivashina2015dollar,dal2012short,eisenschmidt2018measuring} \\
\midrule
US Stock Index & 26 & Dow Jones Industrial Average & \cite{pan2007dynamic,tsai2019predict,pandey2018review,nieh2001dynamic,lin2012comovement,phylaktis2005stock,inci2014dynamic,moore2014dynamic,tsai2012relationship} \\
 & 27 & S\&P 500 & \\
\midrule
EU Stock Index & 28 & Euro Stoxx 50 & \cite{pan2007dynamic,tsai2019predict,pandey2018review,nieh2001dynamic,lin2012comovement,phylaktis2005stock,inci2014dynamic,moore2014dynamic,tsai2012relationship} \\
 & 29 & STOXX 600 & \\
\midrule
US Stock Index Futures & 30 & Dow Jones Futures - Jun & \cite{tah2021dynamic,agrawal2010study,andreou2013stock} \\
\midrule
EU Stock Index Futures & 31 & EURO STOXX 50 Futures - Jun & \cite{tah2021dynamic,agrawal2010study,andreou2013stock} \\
\midrule
CBOE Volatility Index & 32 & VIX & \cite{brunnermeier2008carry,cairns2007exchange,pan2019improving} \\
\bottomrule
\end{tabular}
}
\end{table*}

All financial indicators used in this study can be found in the Table \ref{tab:financial_indicators}. These indicators are collected from the same online platform within the same period.

\subsection{Sentiment and Content Categorization}

Our research distinguishes the collected textual data into two primary categories: \textit{news} and \textit{sentiment analysis}. News texts encompass a broad range of topics, such as political events, economic data releases, and central bank decisions, all of which can influence currency trends. Sentiment-focused texts, however, are primarily analytical, often featuring technical analysis, market predictions, and investment strategies, thus exhibiting a more objective and rational tone. This categorization enables tailored approaches in data processing and prediction tasks based on the distinct characteristics of each text type.

\subsection{Text Preprocessing and Annotation}

Further preprocessing of the raw textual data involves techniques such as tokenization, where elements like stop words, punctuation, and non-alphanumeric characters are eliminated. Commonly used stop words, like 'the', 'is', and 'at', are removed as they do not contribute meaningful information. This cleansing step primes the data for thematic language analysis, ensuring that only significant information remains. In the realm of large language models, effective data processing requires specific prompt words to guide tasks such as text cleansing and annotation. Utilizing the API, we introduce prompt engineering to refine and optimize text processing for sentiment and content annotation. This approach removes irrelevant material, particularly unrelated currency data, to focus exclusively on EUR/USD content. A detailed description of prompt words and techniques used is available in Appendix A and B.

\subsection{Sentiment Scoring and Fine-tuning RoBERTa-Large}

Following the cleansing stage, sentiment intensity within the textual data is assessed using the gpt-4-1106-preview API. Each text is assigned an emotional intensity score ranging from \([-1.0, 1.0]\), where scores at or below \(-1.0\) indicate negative sentiment, and scores at or above \(1.0\) indicate positive sentiment, with values around zero representing neutrality. This scoring provides a quantitative basis for analyzing the emotional tone's impact on exchange rate movements.

To adapt RoBERTa-Large for our currency-focused sentiment analysis and pattern detection, we fine-tune the open-source model from Hugging Face on our annotated dataset. This fine-tuning involves multiple training iterations, during which the model's internal parameters are adjusted to better capture the linguistic nuances of exchange rate discourse and discern unique patterns associated with financial language. This customization of RoBERTa-Large allows it to accurately analyze sentiment and identify underlying trends within the specific context of EUR/USD exchange rate texts, enhancing the model's predictive capacity compared to traditional CNN-based approaches.

\subsection{Sentiment Index Construction}

Recognizing the impact of news on market sentiment over time, we develop a sentiment index (SI) inspired by the work of Xu and Berkely (2014). This SI model assumes that the influence of news decays exponentially, with the most significant impact occurring within the first seven days post-publication. The sentiment index \( SI = e^{-m/7} \) models this decay, where \( m \) represents the number of days since the news release. The sentiment intensity on a given day \( t \), denoted as \( SI_t \), is calculated by summing the Sentiment Value (SV) of that day with the SVs from preceding days, weighted by an exponential decay factor \( e^{-(t-i)/7} \) to account for the influence of past sentiments:

\begin{equation}
SI_t = \sum_{i=1}^{t-1} e^{-\frac{t-i}{7}} SV_i + SV_t,
\end{equation}

This approach more accurately reflects the persistence of news impact on the EUR/USD exchange rate, enhancing the predictive power of our sentiment analysis.

\subsection{Topic Scores for Polarity, Subjectivity, and RoBERTa Analysis}

We calculate topic scores for polarity, subjectivity, and RoBERTa analysis as follows:

\begin{equation}
P_{k,t} = \frac{1}{n_{k,t}} \sum P_i,
\end{equation}
where \( P_{k,t} \) denotes the polarity-topic score of topic \( k \) on day \( t \), and \( n_{k,t} \) is the number of polarity instances for that topic on that day.

\begin{equation}
S_{k,t} = \frac{1}{n_{k,t}} \sum S_i,
\end{equation}
where \( S_{k,t} \) represents the subjectivity-topic score of topic \( k \) on day \( t \).

\begin{equation}
C_{k,t} = \frac{1}{n_{k,t}} \sum C_i,
\end{equation}
where \( C_{k,t} \) denotes the RoBERTa-topic score of topic \( k \) on day \( t \).

\subsection{Lag Order Selection and Feature Selection}

The selection of lags in our study employs the Vector Autoregression (VAR) method. This approach is crucial for determining the optimal number of lags to be used within the model, as an appropriate lag length can more accurately predict exchange rate movements. Initially, we assess the feasibility of different lag lengths. For each potential lag length, a separate VAR model is estimated, and its effectiveness is evaluated using criteria such as the Akaike Information Criterion (AIC) and the Bayesian Information Criterion (BIC). These evaluation standards help us identify the lag length that best captures the temporal dynamics in the data while avoiding overfitting. Through this method, we determine the most suitable lag length for our dataset, providing a solid foundation for the subsequent exchange rate prediction model.

\begin{table*}[!ht]
\caption{Lag order selection result.}
\label{table: lag}
\centering
\resizebox{\textwidth}{!}{%
\begin{tabular}{l c c l c c}
\toprule
\textbf{Variable} & \textbf{Lag} & \textbf{AIC} & \textbf{Variable} & \textbf{Lag} & \textbf{AIC} \\ 
\midrule
USDMXN Exchange Rate        & 2 & -13.46* & Gold Futures                    & 1 & -13.06* \\ 
USDCNY Exchange Rate        & 1 & -13.82* & Copper Futures                  & 2 & -13.26* \\ 
USDJPY Exchange Rate        & 1 & -13.16* & Corn Futures                    & 4 & -13.20* \\ 
USDKRW Exchange Rate        & 1 & -13.28* & Soybeans Futures                & 1 & -12.93* \\ 
EURCNY Exchange Rate        & 2 & -13.74* & US 10-Year Bond Yield           & 1 & -13.41* \\ 
EURGBP Exchange Rate        & 2 & -12.42* & Eurozone 10-Year Gov Bond Yield & 1 & -13.52* \\ 
EURCHF Exchange Rate        & 1 & -13.29* & Euro Stoxx 50 (STOXX50E)        & 2 & -13.40* \\ 
EURRUB Exchange Rate        & 1 & -13.58* & STOXX 600 (STOXX)               & 2 & -12.99* \\ 
EURTRY Exchange Rate        & 1 & -15.29* & Dow Jones Industrial Average (DJI) & 1 & -13.52* \\ 
US Dollar Index             & 2 & -13.51* & S\&P 500 (SPX)                   & 3 & -13.35* \\ 
Euro Index                  & 6 & -14.06* & EURO STOXX 50 (FESX) Futures - Jun & 3 & -13.51* \\ 
US Dollar Futures - Jun     & 2 & -13.47* & Dow Jones Futures - Jun          & 2 & -13.53* \\ 
Euro Futures - Jun          & 2 & -13.57* & Chicago Board Options Exchange Volatility Index & 1 & -12.32* \\ 
Crude Oil WTI Futures       & 7 & -12.96* & CBOE EuroCurrency Volatility Index & 2 & -12.44* \\ 
Natural Gas Futures         & 1 & -13.06* & KBW Nasdaq Bank Index            & 1 & -13.39* \\ 
\bottomrule
\end{tabular}%
}
\end{table*}

For feature selection, our study utilizes the Recursive Feature Elimination (RFE) method, based on the Random Forest regression approach. The RFE method starts with the full set of features and iteratively removes the least important feature at each step. In each iteration, the model is retrained, and the significance of each feature is assessed, typically based on the feature's importance scores in decision trees or its contribution to model performance. This process continues until the predetermined number of features is reached or further removal of features leads to a significant decrease in model performance. The RFE method enables us to effectively identify and select those features that have the most substantial impact on EUR/USD exchange rate prediction, thereby enhancing the model's accuracy and interpretability.

\subsection{Data Segmentation}

In this study, our dataset is strategically segmented into two distinct phases for analysis. The initial phase is the training period, spanning 1520 days, beginning from February 11, 2017, and concluding on April 4, 2022. During this period, the RoBERTa-Large model is utilized for conducting sentiment analysis and identifying hidden patterns within the data. Following the training phase, we proceed to the forecasting period, which covers the next 450 days from April 4, 2022, to June 17, 2023. During this phase, RoBERTa-Large continues to be employed for both sentiment analysis and pattern detection, forming the foundation for forecasting the EUR/USD exchange rate. Additionally, the dataset includes a critical span of 300 days preceding the forecasting period to enhance model robustness. The forecasting phase itself extends for 155 days, culminating on January 19, 2024. This temporal division, with a substantial training duration followed by a focused forecasting period, facilitates comprehensive model training and ensures strong predictive performance in the analysis of the EUR/USD exchange rate.

\begin{figure*}[!ht]
    \centering
    \includegraphics[width=0.8\linewidth]{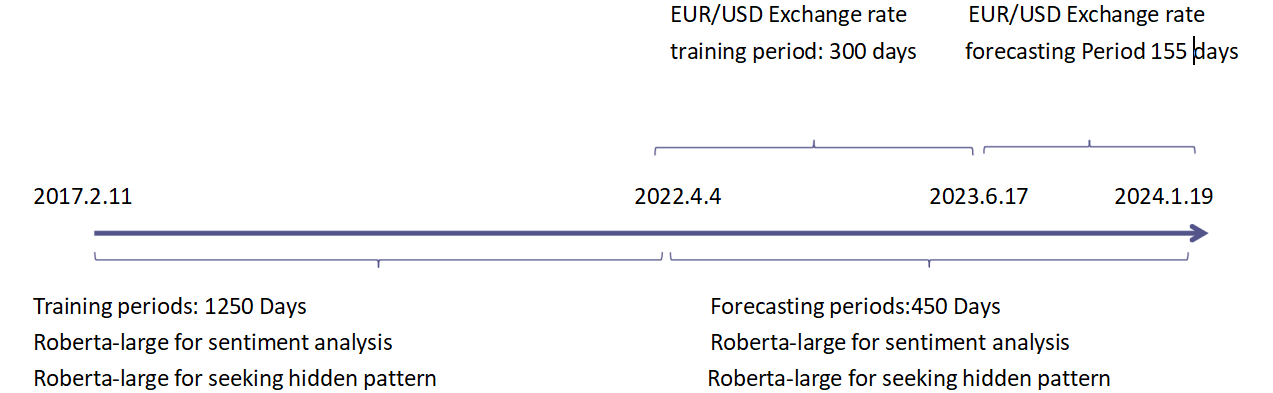}
    \caption{Timeline of training and forecasting periods for EUR/USD exchange rate analysis.}
    \label{fig:timeline}
\end{figure*}

\subsection{Topic Mining and Trend Analysis Results}

\begin{figure*}[!ht]
    \centering
    \includegraphics[width=1\linewidth]{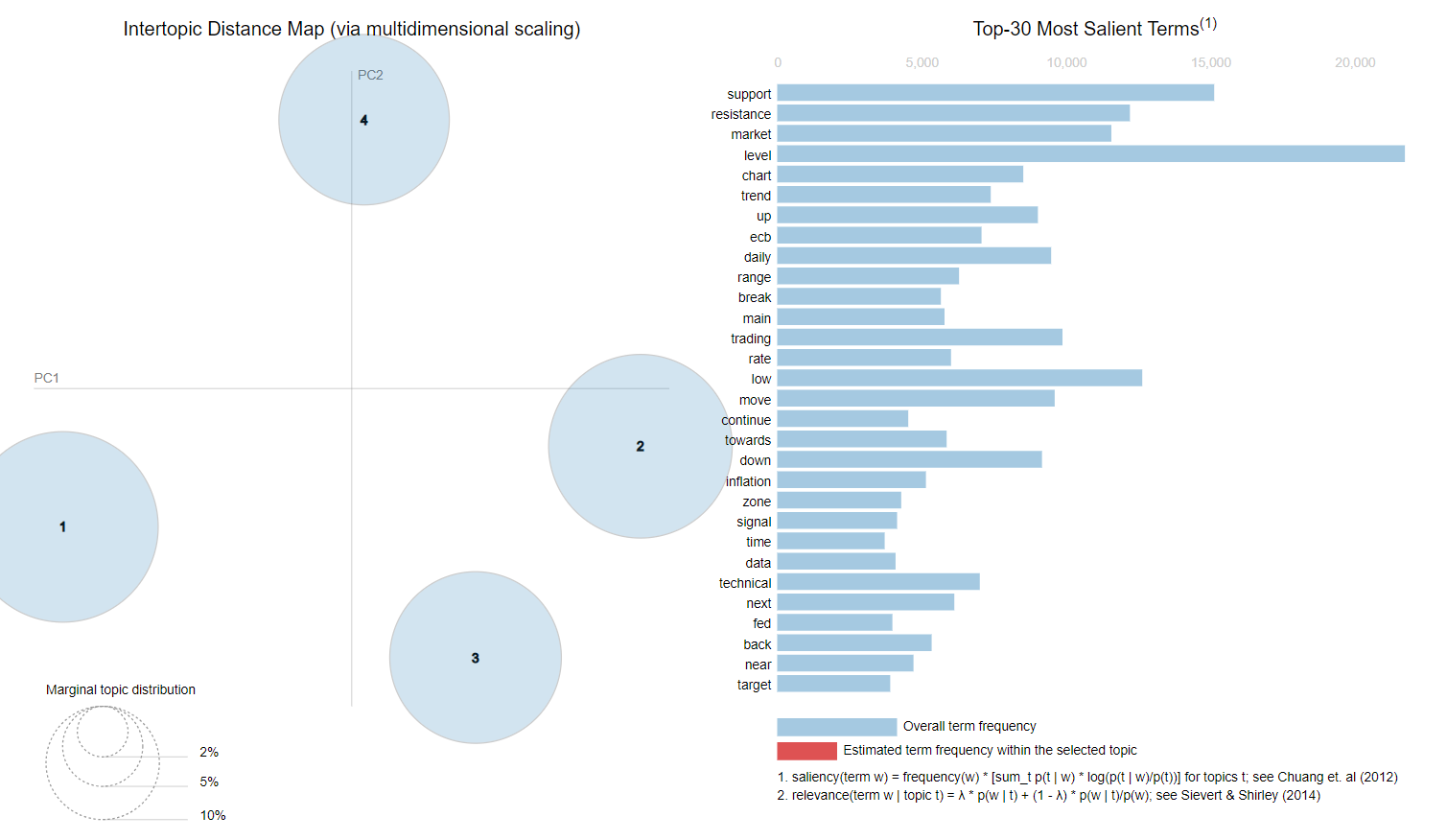}
    \caption{Visualization result of LDA.}
    \label{fig:lda_result}
\end{figure*}

The Figure \ref{fig:lda_result} shows that the best topic of texts for this study is four, and from the Table \ref{table:lda_keywords} the topic division is reasonable, to be more specific: 

The topic of \textit{Economic Fundamentals and Monetary Policy} primarily encompasses policies of the European Central Bank (ECB) and the Federal Reserve (Fed), along with fundamental factors such as inflation, economic data, and interest rate decisions. Keywords include \textit{ECB}, \textit{inflation}, \textit{Fed}, \textit{interest}, \textit{policy}, \textit{economic}, and \textit{data}. Compared to previous discussions, this topic shows an increased frequency of specific economic indicators such as the Purchasing Managers' Index (PMI) and the Consumer Price Index (CPI).

The topic of \textit{Technical Analysis and Key Price Levels} pertains to the technical analysis of currency exchange trends, with a particular emphasis on critical price levels, such as support and resistance. It also encompasses technical indicators like moving averages and momentum. Keywords for this topic include \textit{support}, \textit{resistance}, \textit{level}, \textit{moving average}, and \textit{momentum}. This topic shifts focus toward price levels, with decreased attention to trends and retracements compared to earlier analyses.

\textit{Market Sentiment and Trading Expectations} reflects the moods and future outlook of market participants, including assessments on trend continuation and potential turning points. Keywords here include \textit{market}, \textit{continue}, \textit{break}, \textit{term}, and \textit{forecast}. Relative to prior topics, this one more prominently displays traders' expectations and judgments on future market movements.

The \textit{Price Trends and Comparative Strength of Bulls and Bears} topic focuses on analyzing the specific characteristics of price movements and the resultant interplay of bullish and bearish forces. This includes various types of trends such as ranges, breakouts, and reversals, as well as the ebb and flow of strength between buyers (bulls) and sellers (bears). Keywords include \textit{trend}, \textit{range}, \textit{breakout}, \textit{bulls}, and \textit{bears}. This topic provides a more comprehensive depiction of the various features of price movements, highlighting the strategic plays between opposing market forces.

\begin{table*}[!ht]
\caption{LDA keywords distribution and weights.}
\label{table:lda_keywords}
\centering
\resizebox{\textwidth}{!}{%
\begin{tabular}{|c|c|c|c|c|c|c|c|c|c|c|c|}
\hline
\textbf{No.} & \textbf{Keyword} & \textbf{Weight} & \textbf{No.} & \textbf{Keyword} & \textbf{Weight} & \textbf{No.} & \textbf{Keyword} & \textbf{Weight} & \textbf{No.} & \textbf{Keyword} & \textbf{Weight} \\ \hline
1 & ecb       & 0.019 & 2 & support    & 0.049 & 3 & market    & 0.038 & 4 & chart     & 0.024 \\ \hline
1 & rate      & 0.016 & 2 & resistance & 0.04  & 3 & level     & 0.034 & 4 & up        & 0.024 \\ \hline
1 & inflation & 0.014 & 2 & level      & 0.037 & 3 & day       & 0.019 & 4 & daily     & 0.023 \\ \hline
1 & european  & 0.011 & 2 & towards    & 0.017 & 3 & break     & 0.019 & 4 & low       & 0.022 \\ \hline
1 & data      & 0.011 & 2 & low        & 0.014 & 3 & continue  & 0.015 & 4 & trend     & 0.021 \\ \hline
1 & fed       & 0.011 & 2 & next       & 0.014 & 3 & back      & 0.013 & 4 & trading   & 0.02  \\ \hline
1 & us        & 0.011 & 2 & near       & 0.013 & 3 & us        & 0.012 & 4 & move      & 0.019 \\ \hline
1 & bank      & 0.01  & 2 & price      & 0.012 & 3 & time      & 0.012 & 4 & high      & 0.018 \\ \hline
1 & expected  & 0.01  & 2 & moving     & 0.01  & 3 & term      & 0.011 & 4 & range     & 0.018 \\ \hline
1 & central   & 0.009 & 2 & bearish    & 0.01  & 3 & session   & 0.011 & 4 & down      & 0.017 \\ \hline
\end{tabular}%
}

\end{table*}

Overall, this thematic division aligns fundamentally with previous approaches but is optimized in several aspects:
\begin{itemize}
    \item The first theme now more prominently features specific economic indicators, rendering fundamental analysis more tangible.
    \item The second theme concentrates on key price levels, accentuating the focal points of technical analysis.
    \item The third theme more vividly captures the market participants' predictions for the future, reflecting market sentiment and expectations.
    \item The fourth theme offers a more detailed and comprehensive portrayal of the characteristics of price movements, while also highlighting the dynamics of the contest between bulls and bears.
\end{itemize}

\begin{figure}[!ht]
    \centering
    \includegraphics[width=1\linewidth]{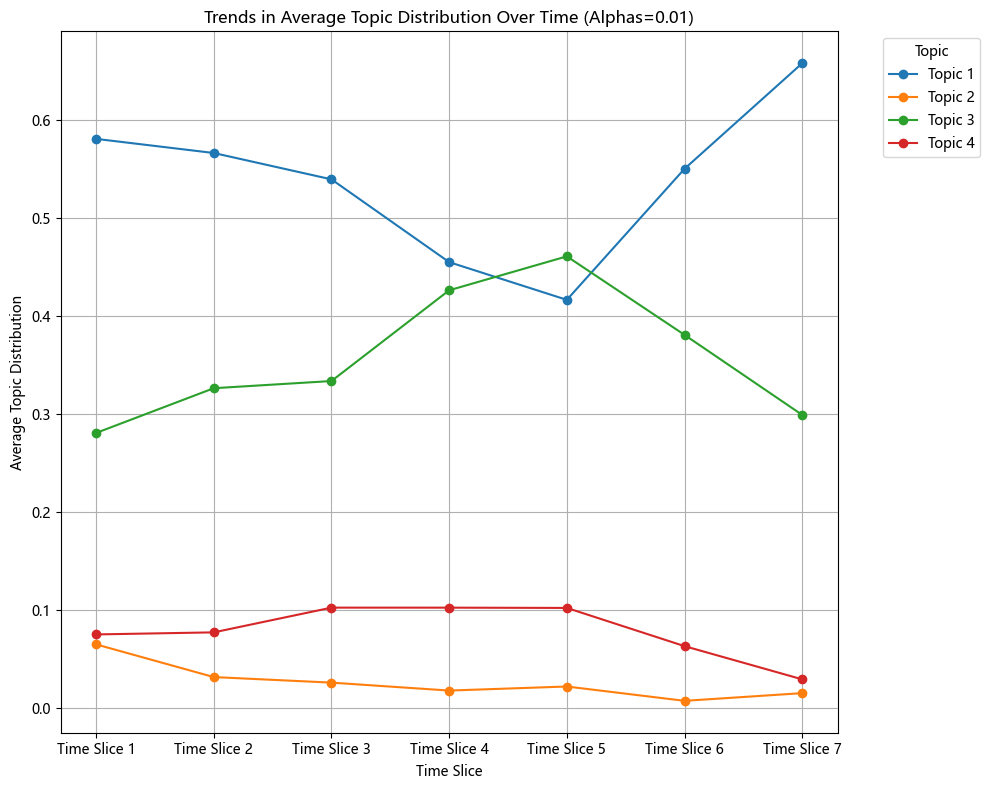}
    \caption{Trend analysis.}
    \label{fig:trend}
\end{figure}

The Figure \ref{fig:trend} shows the trend of topics over the period and the topics over the period can be treated as stable. To be more specific:

\textbf{Topic 1:} This topic starts as the most prominent in the first time slice but shows a steep decline by the second time slice, suggesting that the themes or issues it encompasses may have lost relevance or been overshadowed by other emerging topics. There is a slight uptick in the third time slice before the topic stabilizes and maintains a low profile throughout the remaining periods.

\textbf{Topic 2:} Exhibiting a trend contrary to Topic 1, this topic sees a surge in relevance from the second time slice, overtaking Topic 1. It peaks in the fourth time slice, indicating a period where the subject matter of Topic 2 is of significant interest or concern. Following this peak, there is a marked decline, suggesting a possible resolution to the topic or a shift in focus to other areas.

\textbf{Topic 3:} This topic demonstrates moderate fluctuation but maintains a relatively stable presence throughout all time slices. It doesn't show extreme peaks or troughs, which could imply a consistent level of interest or relevance in the themes it covers.

\textbf{Topic 4:} Starting as the least prominent in the first time slice, there is a dramatic increase in Topic 4's prevalence from the third time slice onward, becoming the most dominant topic by the final time slice. This suggests an escalating interest or growing importance of the issues within Topic 4 over time, possibly indicating emerging trends or newly significant matters drawing attention in the later stages.

The changes in topic prominence over time could reflect shifts in public interest, policy changes, market developments, or other external events affecting the subjects associated with these topics. To provide a more detailed analysis, one would need to examine the specific content and thematic elements within each topic, as represented by the most frequent and relevant terms or phrases for each time period.

\subsection{RoBERTa-Large for Sentiment Analysis Results}

\begin{table}[!ht]
\centering
\caption{RoBERTa-Large Model Performance Metrics for Sentiment Analysis}
\begin{tabular}{ccccc}
\toprule
MSE & MAE & RMSE & R\textsuperscript{2} \\ 
\midrule
0.0162 & 0.0911 & 0.1273 & 0.7976 \\ 
\bottomrule
\end{tabular}
\end{table}

For the sentiment analysis, the results indicate the model has a Mean Squared Error (MSE) of 0.016194, a MAE of 0.091122, a RMSE of 0.127255, and an R-squared ($R^2$) value of 0.797603. These metrics suggest that the model has a fairly good fit with the data, explaining around 79.76\% of the variance in the target variable. The RMSE shows the standard deviation of the residuals, which are relatively low, indicating good predictive accuracy.

\subsection{RoBERTa-Large for Classification Results}

\begin{table}[!ht]
\centering
\caption{Class-wise Performance Metrics for Sentiment Classification}
\begin{tabular}{cccc}
\toprule
Class & Precision & Recall & F1 Score \\ 
\midrule
0 & 0.68 & 0.58 & 0.62 \\ 
1 & 0.64 & 0.74 & 0.69 \\ 
macro avg & 0.66 & 0.66 & 0.65 \\ 
weighted avg & 0.66 & 0.66 & 0.66 \\ 
\bottomrule
\end{tabular}
\end{table}

For the classification task, the model exhibits different performance metrics for two classes (labeled as "0" and "1"). For class "0", the precision is 0.68, recall is 0.58, and the F1 score is 0.62. For class "1", the precision is 0.64, recall is 0.74, and the F1 score is 0.69. The macro average, which gives equal weight to each class, is 0.66 for both precision and recall, with an F1 score of 0.65. The weighted average, which accounts for class imbalance, is also 0.66 for precision, recall, and F1 score. These results suggest that the model is better at identifying class "1" than class "0", as indicated by the higher recall and F1 score for class "1".

\subsection{Classification Metrics Formulas}

\begin{equation}
\text{Precision} = \frac{TP}{TP + FP}
\end{equation}

\begin{equation}
\text{Recall} = \frac{TP}{TP + FN}
\end{equation}

\begin{equation}
F1\text{Score} = 2 \times \frac{\text{Precision} \times \text{Recall}}{\text{Precision} + \text{Recall}}
\end{equation}

\noindent
where \( TP \) denotes true positives, \( FP \) denotes false positives, and \( FN \) denotes false negatives.

\section{Empirical Result}

\subsection{Comparison Experiment}

\begin{table*}[!ht]
\centering
\caption{Model Performance Comparison Across Different Forecasting Methods}

\begin{tabular}{llcccc}
\toprule
Method & Model & MAE & Rank & RMSE & Rank \\ 
\midrule
\multirow{2}{*}{Deep Learning + Optimization} & PSO-LSTM & 0.144535 & 1 & 0.095793 & 1 \\ 
& PSO-GRU & 0.171519 & 9 & 0.192560 & 11 \\ 
\midrule
\multirow{2}{*}{Deep Learning} & LSTM & 0.160276 & 7 & 0.107104 & 3 \\ 
& GRU & 0.154488 & 6 & 0.108713 & 4 \\ 
\midrule
\multirow{2}{*}{Machine Learning + Optimization} & PSO-SVM & 0.142365 & 3 & 0.102022 & 2 \\ 
& PSO-SVR & 0.164589 & 8 & 0.183262 & 9 \\ 
\midrule
\multirow{2}{*}{Machine Learning} & SVM & 0.143759 & 4 & 0.177816 & 6 \\ 
& SVR & 0.189177 & 11 & 0.187001 & 10 \\ 
\midrule
\multirow{2}{*}{Statistical Method (Multi-series)} & VAR & 0.196366 & 12 & 0.197862 & 12 \\ 
& ECM & 0.138344 & 2 & 0.179916 & 8 \\ 
\midrule
\multirow{2}{*}{Statistical Method (Single-series)} & ARIMA & 0.179173 & 10 & 0.146148 & 5 \\ 
& GRACH & 0.152889 & 5 & 0.178053 & 7 \\ 
\bottomrule
\end{tabular}%
\label{fig:A}
\end{table*}

The comparison experiment outlined in the Table \ref{fig:A} presents a comprehensive evaluation of various forecasting models, categorized into three primary methodologies: Deep Learning with optimization, standalone Deep Learning, and traditional Machine Learning with and without optimization, as well as classical Statistical Methods.

In the realm of Deep Learning enhanced by optimization techniques, the PSO-LSTM model demonstrates superior performance, achieving the lowest MAE and Root Mean Square Error (RMSE) across all models tested, justifying its first-place ranking and a weighted rank of 1. The PSO-GRU model, however, lagged significantly behind, ranking ninth in MAE and eleventh in RMSE, culminating in a weighted rank of 10, which indicates a need for further parameter tuning or a reevaluation of its optimization process.

When examining standalone Deep Learning models, the LSTM and GRU variants performed commendably, securing third and fourth places in RMSE, respectively. Their performance highlights the efficacy of deep learning architectures in capturing the sequential dependencies present in financial time series data without the need for external optimization.

Within the Machine Learning domain, the addition of optimization, as seen in the PSO-SVM model, resulted in a notable improvement in MAE, earning it a third-place ranking. However, the PSO-SVR model's performance was less optimal, underscoring the varying effectiveness of optimization depending on the underlying model structure.

The performance of traditional Machine Learning models, SVM and SVR, illustrates their robustness with the SVM model outperforming its optimized counterpart in RMSE and securing a sixth-place rank, emphasizing the importance of model selection before optimization.

The Statistical Methods showcased a mix of results with the VAR model performing the weakest, likely due to its inability to capture the non-linear patterns present in the financial data, as reflected in its twelfth-place ranking for both MAE and RMSE. In contrast, the ECM model's second-place rank for MAE showcases the potential of multi-series methods in capturing complex market relationships.

In the single-series statistical approach, the ARIMA model presented a middling performance, whereas the GRACH model, despite its fifth-place rank for MAE, suggests its relevance in capturing volatility, a critical aspect of exchange rate movements.

The weighted rank, a composite measure based on MAE and RMSE rankings, provides an overarching assessment of model performance, guiding stakeholders in model selection based on a balance of accuracy and predictive reliability.

This comparative experiment elucidates the trade-offs between model complexity, optimization, and forecasting efficacy, offering valuable insights for future model development and selection in the field of financial forecasting.

\subsection{Experiment with and without Textual Data (Ablation Study)}

\begin{table*}[!ht]
\centering
\caption{Impact of Textual and Financial Data on Forecasting Accuracy}

\begin{tabular}[!ht]{lccccc}
\toprule
& Model & Text Features & Financial Features & Combination & Percentage Improvement \\ 
\midrule
\multirow{4}{*}{MAE} 
& PSO-LSTM & 0.0590 & 0.0903 & 0.0746 & 17.2946 \\ 
& LSTM & 0.0594 & 0.0481 & 0.0488 & -1.3134 \\ 
& VAR & 0.0744 & 0.0681 & 0.0712 & -4.5769 \\ 
& Linear Regression & 0.0642 & 0.0494 & 0.0568 & -15.0516 \\ 
\midrule
\multirow{4}{*}{RMSE} 
& PSO-LSTM & 0.0967 & 0.0611 & 0.0327 & 46.4677 \\ 
& LSTM & 0.0445 & 0.0933 & 0.0689 & 26.1479 \\ 
& VAR & 0.0813 & 0.0164 & 0.0488 & -197.8315 \\ 
& Linear Regression & 0.0576 & 0.0178 & 0.0377 & -111.4217 \\ 
\bottomrule
\end{tabular}%
\end{table*}

Using the MAE and Root Mean Square Error (RMSE) as metrics, the study assesses the performance of models like PSO-LSTM, LSTM, VAR, and linear regression with three types of data inputs: text features, financial futures, and a combination of both.

From the textual analysis, we observe the following formulas used for calculating the percentage improvement when incorporating both text and financial futures data, compared to using only financial futures:

\begin{equation}
\text{Improvement Rate (MAE)} = \frac{\text{MAE}_{\text{Financial}} - \text{MAE}_{\text{Combined}}}{\text{MAE}_{\text{Financial}}}
\end{equation}

and

\begin{equation}
\text{Improvement Rate (RMSE)} = \frac{\text{RMSE}_{\text{Financial}} - \text{RMSE}_{\text{Combined}}}{\text{RMSE}_{\text{Financial}}}
\end{equation}

These formulas quantify the performance change when adding text features. Positive improvement rates for \textbf{PSO-LSTM} and \textbf{LSTM} indicate enhanced model performance with text features, with \textbf{PSO-LSTM} achieving around a 17.3\% increase in MAE and a 46.5\% increase in RMSE. In contrast, negative values for \textbf{VAR} and \textbf{Linear Regression} suggest a performance decrease, likely due to the models’ limited suitability for text data.

The findings underscore the importance of feature selection and data fusion in the realm of predictive modeling, highlighting that the synergistic use of quantitative and qualitative data can lead to substantial improvements in the forecasting capabilities of certain models. However, the negative percentages also suggest that careful consideration must be given to the nature of the models and the data they are being trained on, as indiscriminate addition of features can degrade model performance. This nuanced understanding of data integration paves the way for further research into model-specific feature engineering and the potential for custom-tailored data enrichment techniques. 

\subsection{Ablation Experiment Based on PSO-LSTM}

In this part, we have three kinds of textual data. Kind 1: news sentiment score and analysis sentiment score; Kind 2: classification-news and classification analysis; Kind 3: LDA1 and LDA2.

\begin{table}[!ht]
\centering
\caption{Ablation Study Results Based on PSO-LSTM Model with Various Textual Data Inputs}
\begin{tabular}{lcccc}
\toprule
\textbf{Textual Data Combination} & MAE & Rank & RMSE & Rank \\ 
\midrule
Full textual data & 0.1445 & 1 & 0.0958 & 1 \\ 
Kind 1 + Kind 2 & 0.4003 & 3 & 0.4567 & 3 \\ 
Kind 2 + Kind 3 & 0.2983 & 6 & 0.4167 & 6 \\ 
Kind 1 + Kind 3 & 0.3407 & 5 & 0.2534 & 5 \\ 
Kind 1 & 0.3610 & 4 & 0.4855 & 4 \\ 
Kind 2 & 0.3421 & 2 & 0.2750 & 2 \\ 
Kind 3 & 0.3761 & 7 & 0.3116 & 7 \\ 
\bottomrule
\end{tabular}%
\end{table}

This ablation study demonstrates the impact of different combinations of textual data on the PSO-LSTM model's accuracy. The "Full textual data" configuration, encompassing all types of data, achieved the highest performance, indicating that a comprehensive dataset of market sentiment and topic relevance provides significant context for forecasting.

\subsection{DM Test with Different Optimization Algorithms}

\begin{table}[!ht]
\centering
\caption{Ranking of Forecasting Models with Different Optimization Algorithms in DM Test}
\begin{tabular}{lclc}
\toprule
Model & Rank & Model & Rank \\ 
\midrule
PSO-LSTM & 1 & SVR & 2 \\ 
PSO-GRU & 13 & VAR & 8 \\ 
LSTM & 7 & ECM & 12 \\ 
GRU & 3 & CS-LSTM & 4 \\ 
PSO-SVM & 10 & WOA-LSTM & 5 \\ 
PS-SVR & 11 & GA-LSTM & 14 \\ 
SVM & 9 & BAT-LSTM & 6 \\ 
\bottomrule
\end{tabular}%
\end{table}

The optimization algorithms featured in this study include Cuckoo Search (CS) \cite{yang2009cuckoo}, Whale Optimization Algorithm (WOA) \cite{mirjalili2016whale}, Genetic Algorithm (GA) \cite{goldberg1988genetic}, and Bat Algorithm (BAT) \cite{yang2010new}, each applied to LSTM models, hence the notations CS-LSTM, WOA-LSTM, GA-LSTM, and BAT-LSTM. These algorithms are chosen for their ability to navigate complex solution spaces efficiently and are evaluated against models optimized with PPSO \cite{kennedy1995particle} applied to LSTM and GRU networks, as well as Support Vector Machines (SVMs) and Support Vector Regression (SVRs) \cite{cortes1995support}.

The DM test results are intriguing, revealing the PSO-LSTM model as the most accurate, earning the top rank for its forecasting prowess. This highlights the potency of PSO as an optimization technique when combined with the LSTM architecture.

\section{Conclusion}

In conclusion, this research advances EUR/USD exchange rate forecasting by combining deep learning, textual analysis, and PSO. Our approach integrates interdisciplinary techniques, drawing from computer science, finance, and economics to address financial forecasting challenges.

Empirical results highlight the significant advantage of incorporating textual data. By leveraging online news and analysis, the PSO-LSTM model achieved notable accuracy improvements, showcasing the potential of qualitative data to enhance predictive analytics.

Comparative analysis reveals the superior performance of the PSO-LSTM network over classical models, including SVM, SVR, and statistical methods like ARIMA and GARCH. This is supported by the model's top performance across metrics such as MAE and RMSE, further validated by the Diebold-Mariano test.

Through ablation studies, we examined the impact of various textual data categories, including sentiment scores, classification outputs, and LDA topics. Results indicate that comprehensive integration of these data types leads to the most effective forecasting performance.

\section{Future Research}

This study’s reliance on data from limited online sources highlights an area for future improvement. Expanding to include a broader array of platforms, social media, and potentially non-English content would provide a more comprehensive understanding of market sentiment and reveal additional predictors of exchange rate fluctuations. Exploring other optimization algorithms, such as GA or ACO, could further refine the PSO-LSTM model's performance. Additionally, ensemble methods combining forecasts from multiple models may offer more robust and resilient predictions.

Enhanced sentiment analysis is another promising direction. Future research could examine a broader range of emotions beyond polarity, such as trust and anticipation, which may hold predictive value for market trends. Adapting newer machine learning models, like Transformer-based architectures, for time-series forecasting is another rich area for exploration. These models have shown great success in natural language processing and may offer improvements in financial forecasting. Finally, real-time forecasting systems capable of processing live data streams could provide valuable insights for high-frequency trading. Addressing challenges of data velocity and accuracy in real-time systems remains a promising avenue for future research.

Overall, these directions hold potential to further revolutionize financial forecasting, contributing to both academic literature and practical applications in market analysis.

\appendix
\section{appendix}
\subsection{Index correlation analysis}
\begin{figure*}[!ht]
    \centering
    \includegraphics[width=1\linewidth]{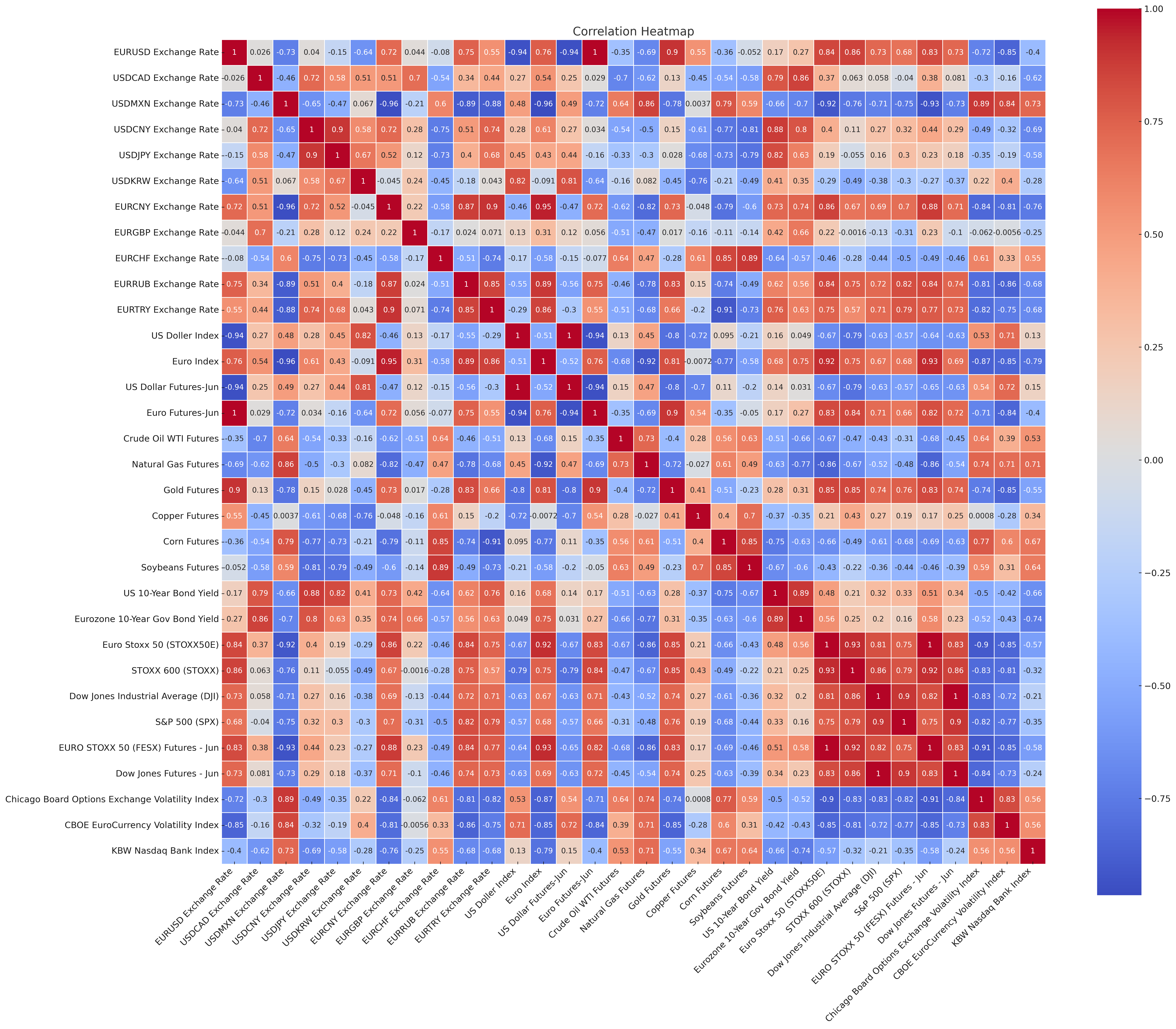}
    \caption{Financial market variables correlation heatmap.}
    \label{fig:heatmap}
\end{figure*}

The Figure \ref{fig:heatmap} shows the correlation analysis of each index within the indicator system. This is essential for establishing the predictive validity and reliability of the model, which allows for the identification of statistically significant predictors, the reduction of multicollinearity, and the enhancement of model robustness, thus ensuring that the forecast of the EUR/USD exchange rate is grounded in empirically verifiable relationships.

\bibliographystyle{IEEEtran}

\bibliography{references}

\begin{thebibliography}{100}
\providecommand{\url}[1]{#1}
\csname url@samestyle\endcsname
\providecommand{\newblock}{\relax}
\providecommand{\bibinfo}[2]{#2}
\providecommand{\BIBentrySTDinterwordspacing}{\spaceskip=0pt\relax}
\providecommand{\BIBentryALTinterwordstretchfactor}{4}
\providecommand{\BIBentryALTinterwordspacing}{\spaceskip=\fontdimen2\font plus
\BIBentryALTinterwordstretchfactor\fontdimen3\font minus
  \fontdimen4\font\relax}
\providecommand{\BIBforeignlanguage}[2]{{%
\expandafter\ifx\csname l@#1\endcsname\relax
\typeout{** WARNING: IEEEtran.bst: No hyphenation pattern has been}%
\typeout{** loaded for the language `#1'. Using the pattern for}%
\typeout{** the default language instead.}%
\else
\language=\csname l@#1\endcsname
\fi
#2}}
\providecommand{\BIBdecl}{\relax}
\BIBdecl

\bibitem{boyoukliev2022modelling}
I.~Boyoukliev, H.~Kulina, and S.~Gocheva-Ilieva, ``Modelling and forecasting of
  eur/usd exchange rate using ensemble learning approach,'' \emph{Cybernetics
  and Information Technologies}, vol.~22, pp. 142--151, 2022.

\bibitem{tsuji2022exchange}
C.~Tsuji, ``Exchange rate forecasting via a machine learning approach,''
  \emph{iBusiness}, 2022.

\bibitem{sarmas2022comparison}
E.~Sarmas, T.~Koutsellis, C.~Ververidis, T.~Papapolyzos, S.~Choumas,
  A.~Bitsikas, and H.~Doukas, ``Comparison of machine learning classifiers for
  exchange rate trend forecasting,'' in \emph{2022 13th International
  Conference on Information, Intelligence, Systems \& Applications (IISA)},
  2022, pp. 1--7.

\bibitem{bejger2021forecasting}
S.~Bejger and P.~Fiszeder, ``Forecasting currency covariances using machine
  learning tree-based algorithms with low and high prices,'' \emph{Przegląd
  Statystyczny}, 2021.

\bibitem{nemavhola2021application}
A.~Nemavhola, C.~Chibaya, and N.~Ochara, ``Application of the lstm - deep
  neural networks - in forecasting foreign currency exchange rates,'' in
  \emph{2021 3rd International Multidisciplinary Information Technology and
  Engineering Conference (IMITEC)}, 2021, pp. 1--6.

\bibitem{kaushik2020forecasting}
M.~Kaushik and A.~K. Giri, ``Forecasting foreign exchange rate: A multivariate
  comparative analysis between traditional econometric, contemporary machine
  learning \& deep learning techniques,'' \emph{ArXiv}, vol. abs/2002.10247,
  2020.

\bibitem{sun2020ensemble}
S.~Sun, S.~Wang, and Y.~Wei, ``A new ensemble deep learning approach for
  exchange rates forecasting and trading,'' \emph{Adv. Eng. Informatics},
  vol.~46, p. 101160, 2020.

\bibitem{escudero2021neural}
P.~Escudero, W.~Alcocer, and J.~Paredes, ``Recurrent neural networks and arima
  models for euro/dollar exchange rate forecasting,'' \emph{Applied Sciences},
  2021.

\bibitem{shen2021hybrid}
M.-L. Shen, C.-F. Lee, H.-H. Liu, P.-Y. Chang, and C.-H. Yang, ``An effective
  hybrid approach for forecasting currency exchange rates,''
  \emph{Sustainability}, 2021.

\bibitem{nemavhola2021lstm}
A.~Nemavhola, C.~Chibaya, and N.~Ochara, ``Application of the lstm - deep
  neural networks - in forecasting foreign currency exchange rates,'' in
  \emph{2021 3rd International Multidisciplinary Information Technology and
  Engineering Conference (IMITEC)}, 2021, pp. 1--6.

\bibitem{cao2020deep}
W.~Cao, W.~Zhu, W.~Wang, Y.~Demazeau, and C.~Zhang, ``A deep coupled lstm
  approach for usd/cny exchange rate forecasting,'' \emph{IEEE Intelligent
  Systems}, vol.~35, pp. 43--53, 2020.

\bibitem{meese1983empirical}
R.~A. Meese and K.~Rogoff, ``Empirical exchange rate models of the seventies:
  Do they fit out of sample?'' \emph{Journal of International Economics},
  vol.~14, no. 1-2, pp. 3--24, 1983.

\bibitem{bollerslev1986generalized}
T.~Bollerslev, ``Generalized autoregressive conditional heteroskedasticity,''
  \emph{Journal of Econometrics}, vol.~31, no.~3, pp. 307--327, 1986.

\bibitem{engel2005exchange}
C.~Engel, ``Exchange rate models are not as bad as you think,'' \emph{NBER
  Macroeconomics Annual}, vol.~20, pp. 381--441, 2005.

\bibitem{rossi2013exchange}
B.~Rossi, ``Exchange rate predictability,'' \emph{Journal of economic
  literature}, vol.~51, no.~4, pp. 1063--1119, 2013.

\bibitem{huang2005forecasting}
W.~Huang, Y.~Nakamori, and S.-Y. Wang, ``Forecasting stock market movement
  direction with support vector machine,'' \emph{Computers \& Operations
  Research}, vol.~32, no.~10, pp. 2513--2522, 2005.

\bibitem{krauss2017statistical}
C.~Krauss, X.~A. Do, and N.~Huck, ``Statistical arbitrage in cryptocurrency
  markets,'' \emph{Journal of Risk and Financial Management}, vol.~10, no.~4,
  p.~23, 2017.

\bibitem{yu2007forecasting}
L.~Yu, S.~Wang, and K.~K. Lai, ``Forecasting crude oil price with an emd-based
  neural network ensemble learning paradigm,'' \emph{Energy Economics},
  vol.~30, no.~5, pp. 2623--2635, 2008.

\bibitem{ozturk2016modeling}
S.~{"O}zt{"u}rk, H.~C. Akda{\u{g}}, and F.~Konak, ``Modeling and forecasting
  long memory in exchange rate volatility versus stable and integrated garch
  models,'' \emph{Applied Economics}, vol.~48, no.~28, pp. 2655--2663, 2016.

\bibitem{nassirtoussi2014text}
A.~K. Nassirtoussi, S.~Aghabozorgi, T.~Y. Wah, and D.~C.~L. Ngo, ``Text mining
  for market prediction: A systematic review,'' \emph{Expert Systems with
  Applications}, vol.~41, no.~16, pp. 7653--7670, 2014.

\bibitem{das2007yahoo}
S.~R. Das and M.~Y. Chen, ``Yahoo! for amazon: Sentiment extraction from small
  talk on the web,'' in \emph{AFA 2006 Boston Meetings Paper}, 2007.

\bibitem{tetlock2007giving}
P.~C. Tetlock, ``Giving content to investor sentiment: The role of media in the
  stock market,'' \emph{The Journal of Finance}, vol.~62, no.~3, pp.
  1139--1168, 2007.

\bibitem{hochreiter1997long}
S.~Hochreiter and J.~Schmidhuber, ``Long short-term memory,'' \emph{Neural
  Computation}, vol.~9, no.~8, pp. 1735--1780, 1997.

\bibitem{cho2014learning}
K.~Cho, B.~Van~Merri{"e}nboer, C.~Gulcehre, D.~Bahdanau, F.~Bougares,
  H.~Schwenk, and Y.~Bengio, ``Learning phrase representations using rnn
  encoder-decoder for statistical machine translation,'' \emph{arXiv preprint
  arXiv:1406.1078}, 2014.

\bibitem{lecun2015deep}
Y.~LeCun, Y.~Bengio, and G.~Hinton, ``Deep learning,'' \emph{Nature}, vol. 521,
  no. 7553, pp. 436--444, 2015.

\bibitem{smales2014news}
L.~A. Smales, ``News sentiment and the investor fear gauge,'' \emph{Finance
  Research Letters}, vol.~11, no.~2, pp. 122--130, 2014.

\bibitem{beckmann2015exchange}
J.~Beckmann, R.~Sch{"u}ssler, and R.~Czudaj, ``Exchange rate forecasts and
  expected fundamentals,'' \emph{Journal of Behavioral Finance}, vol.~16,
  no.~4, pp. 344--354, 2015.

\bibitem{devlin2018bert}
J.~Devlin, M.-W. Chang, K.~Lee, and K.~Toutanova, ``Bert: Pre-training of deep
  bidirectional transformers for language understanding,'' \emph{arXiv preprint
  arXiv:1810.04805}, 2018.

\bibitem{serrano2019using}
C.~Serrano-Cinca and B.~Guti{'e}rrez-Nieto, ``Using sentiment analysis to
  enhance the measurement of macroeconomic uncertainty,'' \emph{International
  Journal of Forecasting}, vol.~35, no.~2, pp. 655--665, 2019.

\bibitem{spalti2022exploiting}
P.~Sp{"a}lti, D.~Boller, L.~Brandt, S.~Dolfus, H.~Dorn, J.~O. Roth, M.~Jann,
  A.~Smirnov, O.~Berka, P.~Gubelmann \emph{et~al.}, ``Exploiting satellite
  imagery and natural language processing for assessing trade relationships
  among countries,'' \emph{arXiv preprint arXiv:2207.11128}, 2022.

\bibitem{brown2020language}
T.~B. Brown, B.~Mann, N.~Ryder, M.~Subbiah, J.~Kaplan, P.~Dhariwal,
  A.~Neelakantan, P.~Shyam, G.~Sastry, A.~Askell \emph{et~al.}, ``Language
  models are few-shot learners,'' \emph{arXiv preprint arXiv:2005.14165}, 2020.

\bibitem{pornwattanavichai2022bertforex}
A.~Pornwattanavichai, S.~Maneeroj, and S.~Boonsiri, ``Bertforex: Cascading
  model for forex market forecasting using fundamental and technical indicator
  data based on bert,'' \emph{IEEE Access}, vol.~PP, pp. 1--1, 2022.

\bibitem{pfahler2021advanced}
J.~F. Pfahler, ``Exchange rate forecasting with advanced machine learning
  methods,'' \emph{Journal of Risk and Financial Management}, 2021.

\bibitem{tak2022foreign}
A.~L.~C. Tak and R.~Logeswaran, ``Foreign currency exchange market prediction
  using machine learning techniques,'' \emph{2022 IEEE International Conference
  on Distributed Computing and Electrical Circuits and Electronics (ICDCECE)},
  pp. 1--5, 2022.

\bibitem{shang2023datasets}
J.~Shang and S.~Hamori, ``Do large datasets or hybrid integrated models
  outperform simple ones in predicting commodity prices and foreign exchange
  rates?'' \emph{Journal of Risk and Financial Management}, 2023.

\bibitem{lin2022deep}
Y.~Lin, C.-J. Lai, and P.-F. Pai, ``Using deep learning techniques in
  forecasting stock markets by hybrid data with multilingual sentiment
  analysis,'' \emph{Electronics}, 2022.

\bibitem{claveria2022deep}
O.~Claveria, E.~Monte, P.~Sorić, and S.~Torra, ``An application of deep
  learning for exchange rate forecasting,'' \emph{SSRN Electronic Journal},
  2022.

\bibitem{argotty2023novel}
M.~Argotty-Erazo, A.~Blázquez-Zaballos, C.~A. Argoty-Eraso, L.~L.
  Lorente-Leyva, N.~N. Sánchez-Pozo, and D.~H. Peluffo-Ordóñez, ``A novel
  linear-model-based methodology for predicting the directional movement of the
  euro-dollar exchange rate,'' \emph{IEEE Access}, vol.~11, pp.
  67\,249--67\,284, 2023.

\bibitem{Bommasani2021}
R.~Bommasani, D.~Hudson, E.~Adeli, R.~Altman, S.~Arora, S.~von Arx,
  M.~Bernstein, J.~Bohg, A.~Bosselut, E.~Brunskill, and E.~Brynjolfsson, ``On
  the opportunities and risks of foundation models,'' 2021.

\bibitem{Brown2020}
T.~Brown, B.~Mann, N.~Ryder, M.~Subbiah, J.~Kaplan, P.~Dhariwal,
  A.~Neelakantan, P.~Shyam, G.~Sastry, A.~Askell, and S.~Agarwal, ``Language
  models are few-shot learners,'' \emph{Advances in Neural Information
  Processing Systems}, vol.~33, pp. 1877--1901, 2020.

\bibitem{Devlin2019}
J.~Devlin, M.~Chang, K.~Lee, and K.~Toutanova, ``Bert: Pre-training of deep
  bidirectional transformers for language understanding,'' \emph{arXiv preprint
  arXiv:1810.04805}, 2019.

\bibitem{Vaswani2017}
A.~Vaswani, N.~Shazeer, N.~Parmar, J.~Uszkoreit, L.~Jones, A.~Gomez, L.~Kaiser,
  and I.~Polosukhin, ``Attention is all you need,'' in \emph{Advances in Neural
  Information Processing Systems}, vol.~30, 2017.

\bibitem{Liu2019}
Y.~Liu, M.~Ott, N.~Goyal, J.~Du, M.~Joshi, D.~Chen, O.~Levy, M.~Lewis,
  L.~Zettlemoyer, and V.~Stoyanov, ``Roberta: A robustly optimized bert
  pretraining approach,'' \emph{arXiv preprint arXiv:1907.11692}, 2019.

\bibitem{Hochreiter1997}
S.~Hochreiter and J.~Schmidhuber, ``Long short-term memory,'' \emph{Neural
  Computation}, vol.~9, no.~8, pp. 1735--1780, 1997.

\bibitem{Graves2012}
A.~Graves, \emph{Supervised sequence labelling with recurrent neural
  networks}.\hskip 1em plus 0.5em minus 0.4em\relax Springer, 2012.

\bibitem{Kennedy1995}
J.~Kennedy and R.~Eberhart, ``Particle swarm optimization,'' in
  \emph{Proceedings of ICNN'95-International Conference on Neural Networks},
  vol.~4.\hskip 1em plus 0.5em minus 0.4em\relax IEEE, 1995, pp. 1942--1948.

\bibitem{Hyndman2006}
R.~Hyndman and A.~Koehler, ``Another look at measures of forecast accuracy,''
  \emph{International Journal of Forecasting}, vol.~22, no.~4, pp. 679--688,
  2006.

\bibitem{radford2019language}
A.~Radford, J.~Wu, R.~Child, D.~Luan, D.~Amodei, and I.~Sutskever, ``Language
  models are unsupervised multitask learners,'' \emph{OpenAI blog}, vol.~1,
  no.~8, p.~9, 2019.

\bibitem{blei2003latent}
D.~M. Blei, A.~Y. Ng, and M.~I. Jordan, ``Latent dirichlet allocation,''
  \emph{Journal of machine Learning research}, vol.~3, no. Jan, pp. 993--1022,
  2003.

\bibitem{blei2006dynamic}
D.~M. Blei and J.~D. Lafferty, ``Dynamic topic models,'' in \emph{Proceedings
  of the 23rd international conference on Machine learning}, 2006, pp.
  113--120.

\bibitem{yang2024weighted}
L.~Yang, H.~Wang, C.~Gu, and H.~Yang, ``Weighted signed networks reveal
  interactions between us foreign exchange rates,'' \emph{Entropy}, vol.~26,
  no.~2, p. 161, 2024.

\bibitem{barunik2017asymmetric}
J.~Barun{\'\i}k, E.~Ko{\v{c}}enda, and L.~V{\'a}cha, ``Asymmetric volatility
  connectedness on the forex market,'' \emph{Journal of International Money and
  Finance}, vol.~77, pp. 39--56, 2017.

\bibitem{greenwood2021measuring}
M.~Greenwood-Nimmo, V.~H. Nguyen, and Y.~Shin, ``Measuring the connectedness of
  the global economy,'' \emph{International Journal of Forecasting}, vol.~37,
  no.~2, pp. 899--919, 2021.

\bibitem{kilic2017contagion}
E.~Kilic, ``Contagion effects of us dollar and chinese yuan in forward and spot
  foreign exchange markets,'' \emph{Economic Modelling}, vol.~62, pp. 51--67,
  2017.

\bibitem{alkan2022currency}
A.~Alkan, A.~F. Alkaya, and P.~Sch{\"u}ller, ``Currency exchange rate
  forecasting with social media sentiment analysis,'' in \emph{Intelligent and
  Fuzzy Techniques for Emerging Conditions and Digital Transformation:
  Proceedings of the INFUS 2021 Conference, held August 24-26, 2021. Volume
  2}.\hskip 1em plus 0.5em minus 0.4em\relax Springer, 2022, pp. 490--497.

\bibitem{ding2021conditional}
L.~Ding, ``Conditional correlation between exchange rates and stock prices,''
  \emph{The Quarterly Review of Economics and Finance}, vol.~80, pp. 452--463,
  2021.

\bibitem{chantarakasemchit2020forex}
O.~Chantarakasemchit, S.~Nuchitprasitchai, and Y.~Nilsiam, ``Forex rates
  prediction on eur/usd with simple moving average technique and financial
  factors,'' in \emph{2020 17th International Conference on Electrical
  Engineering/Electronics, Computer, Telecommunications and Information
  Technology (ECTI-CON)}.\hskip 1em plus 0.5em minus 0.4em\relax IEEE, 2020,
  pp. 771--774.

\bibitem{gurrib2016optimizing}
I.~Gurrib and E.~Elshareif, ``Optimizing the performance of the fractal
  adaptive moving average strategy: The case of eur/usd,'' \emph{International
  Journal of Economics and Finance}, vol.~8, no.~2, pp. 171--179, 2016.

\bibitem{tornell2012speculation}
A.~Tornell and C.~Yuan, ``Speculation and hedging in the currency futures
  markets: Are they informative to the spot exchange rates,'' \emph{Journal of
  Futures Markets}, vol.~32, no.~2, pp. 122--151, 2012.

\bibitem{ferraro2015can}
D.~Ferraro, K.~Rogoff, and B.~Rossi, ``Can oil prices forecast exchange rates?
  an empirical analysis of the relationship between commodity prices and
  exchange rates,'' \emph{Journal of International Money and Finance}, vol.~54,
  pp. 116--141, 2015.

\bibitem{chen2003commodity}
Y.-c. Chen and K.~Rogoff, ``Commodity currencies,'' \emph{Journal of
  international Economics}, vol.~60, no.~1, pp. 133--160, 2003.

\bibitem{lyons1997simultaneous}
R.~K. Lyons, ``A simultaneous trade model of the foreign exchange hot potato,''
  \emph{Journal of international Economics}, vol.~42, no. 3-4, pp. 275--298,
  1997.

\bibitem{cashin2004commodity}
P.~Cashin, L.~F. C{\'e}spedes, and R.~Sahay, ``Commodity currencies and the
  real exchange rate,'' \emph{Journal of Development Economics}, vol.~75,
  no.~1, pp. 239--268, 2004.

\bibitem{jadidzadeh2017does}
A.~Jadidzadeh and A.~Serletis, ``How does the us natural gas market react to
  demand and supply shocks in the crude oil market?'' \emph{Energy Economics},
  vol.~63, pp. 66--74, 2017.

\bibitem{lizardo2010oil}
R.~A. Lizardo and A.~V. Mollick, ``Oil price fluctuations and us dollar
  exchange rates,'' \emph{Energy economics}, vol.~32, no.~2, pp. 399--408,
  2010.

\bibitem{basher2016impact}
S.~A. Basher, A.~A. Haug, and P.~Sadorsky, ``The impact of oil shocks on
  exchange rates: A markov-switching approach,'' \emph{Energy Economics},
  vol.~54, pp. 11--23, 2016.

\bibitem{pukthuanthong2011gold}
K.~Pukthuanthong and R.~Roll, ``Gold and the dollar (and the euro, pound, and
  yen),'' \emph{Journal of Banking \& Finance}, vol.~35, no.~8, pp. 2070--2083,
  2011.

\bibitem{sjaastad2008price}
L.~A. Sjaastad, ``The price of gold and the exchange rates: Once again,''
  \emph{Resources Policy}, vol.~33, no.~2, pp. 118--124, 2008.

\bibitem{zhang2016exchange}
H.~J. Zhang, J.-M. Dufour, and J.~W. Galbraith, ``Exchange rates and commodity
  prices: Measuring causality at multiple horizons,'' \emph{Journal of
  Empirical Finance}, vol.~36, pp. 100--120, 2016.

\bibitem{nazlioglu2012oil}
S.~Nazlioglu and U.~Soytas, ``Oil price, agricultural commodity prices, and the
  dollar: A panel cointegration and causality analysis,'' \emph{Energy
  Economics}, vol.~34, no.~4, pp. 1098--1104, 2012.

\bibitem{akanni2020returns}
L.~O. Akanni, ``Returns and volatility spillover between food prices and
  exchange rate in nigeria,'' \emph{Journal of Agribusiness in Developing and
  Emerging Economies}, vol.~10, no.~3, pp. 307--325, 2020.

\bibitem{nazlioglu2013volatility}
S.~Nazlioglu, C.~Erdem, and U.~Soytas, ``Volatility spillover between oil and
  agricultural commodity markets,'' \emph{Energy Economics}, vol.~36, pp.
  658--665, 2013.

\bibitem{rezitis2015relationship}
A.~N. Rezitis, ``The relationship between agricultural commodity prices, crude
  oil prices and us dollar exchange rates: A panel var approach and causality
  analysis,'' \emph{International Review of Applied Economics}, vol.~29, no.~3,
  pp. 403--434, 2015.

\bibitem{engel2023forecasting}
C.~Engel and S.~P.~Y. Wu, ``Forecasting the us dollar in the 21st century,''
  \emph{Journal of International Economics}, vol. 141, p. 103715, 2023.

\bibitem{chinn2004monetary}
M.~D. Chinn and G.~Meredith, ``Monetary policy and long-horizon uncovered
  interest parity,'' \emph{IMF staff papers}, vol.~51, no.~3, pp. 409--430,
  2004.

\bibitem{lace2015determining}
N.~Lace, I.~Ma{\v{c}}erinskien{\.e}, and A.~Bal{\v{c}}i{\=u}nas, ``Determining
  the eur/usd exchange rate with us and german government bond yields in the
  post-crisis period,'' \emph{Intellectual Economics}, vol.~9, no.~2, pp.
  150--155, 2015.

\bibitem{afonso2018euro}
A.~Afonso and M.~Kazemi, ``Euro area sovereign yields and the power of
  unconventional monetary policy,'' \emph{Finance a {\'u}v{\v{e}}r-Czech
  Journal of Economics and Finance}, vol.~68, no.~2, pp. 100--119, 2018.

\bibitem{cecioni2018ecb}
M.~Cecioni, ``Ecb monetary policy and the euro exchange rate,'' \emph{Bank of
  Italy Temi di Discussione (Working Paper) No}, vol. 1172, 2018.

\bibitem{tonzer2015cross}
L.~Tonzer, ``Cross-border interbank networks, banking risk and contagion,''
  \emph{Journal of Financial Stability}, vol.~18, pp. 19--32, 2015.

\bibitem{ivashina2015dollar}
V.~Ivashina, D.~S. Scharfstein, and J.~C. Stein, ``Dollar funding and the
  lending behavior of global banks,'' \emph{The Quarterly Journal of
  Economics}, vol. 130, no.~3, pp. 1241--1281, 2015.

\bibitem{dal2012short}
M.~Dal~Bianco, M.~Camacho, and G.~P. Quiros, ``Short-run forecasting of the
  euro-dollar exchange rate with economic fundamentals,'' \emph{Journal of
  International Money and Finance}, vol.~31, no.~2, pp. 377--396, 2012.

\bibitem{duffie2015reforming}
D.~Duffie and J.~C. Stein, ``Reforming libor and other financial market
  benchmarks,'' \emph{Journal of Economic Perspectives}, vol.~29, no.~2, pp.
  191--212, 2015.

\bibitem{du2018deviations}
W.~Du, A.~Tepper, and A.~Verdelhan, ``Deviations from covered interest rate
  parity,'' \emph{The Journal of Finance}, vol.~73, no.~3, pp. 915--957, 2018.

\bibitem{eisenschmidt2018measuring}
J.~Eisenschmidt, D.~Kedan, and R.~D. Tietz, ``Measuring fragmentation in the
  euro area unsecured overnight interbank money market,'' \emph{Economic
  Bulletin Articles}, vol.~5, 2018.

\bibitem{pan2007dynamic}
M.-S. Pan, R.~C.-W. Fok, and Y.~A. Liu, ``Dynamic linkages between exchange
  rates and stock prices: Evidence from east asian markets,''
  \emph{International Review of Economics \& Finance}, vol.~16, no.~4, pp.
  503--520, 2007.

\bibitem{tsai2019predict}
Y.-C. Tsai, J.-H. Chen, and J.-J. Wang, ``Predict forex trend via convolutional
  neural networks,'' \emph{Journal of Intelligent Systems}, vol.~29, no.~1, pp.
  941--958, 2019.

\bibitem{pandey2018review}
T.~N. Pandey, A.~K. Jagadev, S.~Dehuri, and S.-B. Cho, ``A review and empirical
  analysis of neural networks based exchange rate prediction,''
  \emph{Intelligent Decision Technologies}, vol.~12, no.~4, pp. 423--439, 2018.

\bibitem{nieh2001dynamic}
C.-C. Nieh and C.-F. Lee, ``Dynamic relationship between stock prices and
  exchange rates for g-7 countries,'' \emph{The Quarterly Review of Economics
  and Finance}, vol.~41, no.~4, pp. 477--490, 2001.

\bibitem{lin2012comovement}
C.-H. Lin, ``The comovement between exchange rates and stock prices in the
  asian emerging markets,'' \emph{International Review of Economics \&
  Finance}, vol.~22, no.~1, pp. 161--172, 2012.

\bibitem{phylaktis2005stock}
K.~Phylaktis and F.~Ravazzolo, ``Stock prices and exchange rate dynamics,''
  \emph{Journal of international Money and Finance}, vol.~24, no.~7, pp.
  1031--1053, 2005.

\bibitem{inci2014dynamic}
A.~C. Inci and B.~S. Lee, ``Dynamic relations between stock returns and
  exchange rate changes,'' \emph{European Financial Management}, vol.~20,
  no.~1, pp. 71--106, 2014.

\bibitem{moore2014dynamic}
T.~Moore and P.~Wang, ``Dynamic linkage between real exchange rates and stock
  prices: Evidence from developed and emerging asian markets,''
  \emph{International Review of Economics \& Finance}, vol.~29, pp. 1--11,
  2014.

\bibitem{tsai2012relationship}
I.-C. Tsai, ``The relationship between stock price index and exchange rate in
  asian markets: A quantile regression approach,'' \emph{Journal of
  International Financial Markets, Institutions and Money}, vol.~22, no.~3, pp.
  609--621, 2012.

\bibitem{tah2021dynamic}
K.~A. Tah and G.~Ngene, ``Dynamic linkages between us and eurodollar interest
  rates: new evidence from causality in quantiles,'' \emph{Journal of Economics
  and Finance}, vol.~45, pp. 200--210, 2021.

\bibitem{agrawal2010study}
G.~Agrawal, A.~K. Srivastav, and A.~Srivastava, ``A study of exchange rates
  movement and stock market volatility,'' \emph{International Journal of
  business and management}, vol.~5, no.~12, p.~62, 2010.

\bibitem{andreou2013stock}
E.~Andreou, M.~Matsi, and A.~Savvides, ``Stock and foreign exchange market
  linkages in emerging economies,'' \emph{Journal of International Financial
  Markets, Institutions and Money}, vol.~27, pp. 248--268, 2013.

\bibitem{brunnermeier2008carry}
M.~K. Brunnermeier, S.~Nagel, and L.~H. Pedersen, ``Carry trades and currency
  crashes,'' \emph{NBER macroeconomics annual}, vol.~23, no.~1, pp. 313--348,
  2008.

\bibitem{cairns2007exchange}
J.~Cairns, C.~Ho, and R.~N. McCauley, ``Exchange rates and global volatility:
  implications for asia-pacific currencies,'' \emph{BIS Quarterly Review,
  March}, 2007.

\bibitem{pan2019improving}
Z.~Pan, Y.~Wang, L.~Liu, and Q.~Wang, ``Improving volatility prediction and
  option valuation using vix information: A volatility spillover garch model,''
  \emph{Journal of Futures Markets}, vol.~39, no.~6, pp. 744--776, 2019.

\bibitem{yang2009cuckoo}
X.-S. Yang and S.~Deb, ``Cuckoo search via l{'e}vy flights,'' \emph{2009 World
  congress on nature \& biologically inspired computing (NaBIC)}, pp. 210--214,
  2009.

\bibitem{mirjalili2016whale}
S.~Mirjalili and A.~Lewis, ``The whale optimization algorithm,'' \emph{Advances
  in engineering software}, vol.~95, pp. 51--67, 2016.

\bibitem{goldberg1988genetic}
D.~E. Goldberg and J.~H. Holland, \emph{Genetic algorithms in search,
  optimization, and machine learning}.\hskip 1em plus 0.5em minus 0.4em\relax
  Addison-wesley Reading, 1988.

\bibitem{yang2010new}
X.-S. Yang, ``A new metaheuristic bat-inspired algorithm,'' \emph{Nature
  inspired cooperative strategies for optimization (NICSO 2010)}, pp. 65--74,
  2010.

\bibitem{kennedy1995particle}
J.~Kennedy and R.~Eberhart, ``Particle swarm optimization,'' in
  \emph{Proceedings of ICNN'95-international conference on neural networks},
  vol.~4.\hskip 1em plus 0.5em minus 0.4em\relax IEEE, 1995, pp. 1942--1948.

\bibitem{cortes1995support}
C.~Cortes and V.~Vapnik, ``Support-vector networks,'' \emph{Machine learning},
  vol.~20, no.~3, pp. 273--297, 1995.

\end{thebibliography}

\end{document}